\documentclass[fleqn,usenatbib]{mnras}

\usepackage{newtxtext,newtxmath}

\usepackage[T1]{fontenc}
\usepackage{multirow}

\DeclareRobustCommand{\VAN}[3]{#2}
\let\VANthebibliography\thebibliography
\def\thebibliography{\DeclareRobustCommand{\VAN}[3]{##3}\VANthebibliography}

\usepackage{graphicx}	
\usepackage{amsmath}	
\usepackage{orcidlink}

\title[Eu/Si in MW Stars and Globular Clusters]{The ratio of [Eu/$\alpha$] differentiates accreted/in-situ Milky Way stars across metallicities, as indicated by both field stars and globular clusters}

\author[S. Monty]{\noindent
Stephanie Monty$^{1}$\thanks{E-mail: sm2744@cam.ac.uk}\orcidlink{0000-0002-9225-5822},
Vasily Belokurov$^{1}$\orcidlink{0000-0002-0038-9584},
Jason L. Sanders$^{2}$\orcidlink{0000-0003-4593-6788},
Terese T. Hansen$^{3,4}$\orcidlink{0000-0001-6154-8983},
\newauthor
Charli M. Sakari$^{5}$\orcidlink{0000-0002-5095-4000},
Madeleine McKenzie$^{6, 7}$\orcidlink{0000-0002-1715-1257},
GyuChul Myeong$^{1}$\orcidlink{0000-0002-5629-8876},
Elliot Y. Davies$^{1}$\orcidlink{0000-0001-5996-4072},
\newauthor
Anke Ardern-Arentsen$^{1}$\orcidlink{0000-0002-0544-2217},
Davide Massari$^{8}$\orcidlink{0000-0001-8892-4301}
\\
$^{1}$ Institute of Astronomy, University of Cambridge, Madingley Rd, Cambridge, CB3 0HA, UK\\
$^{2}$ Department of Physics and Astronomy, University College London, London WC1E 6BT, UK\\
$^{3}$ Department of Astronomy, Stockholm University, AlbaNova University Center, SE-106 91 Stockholm, Sweden \\
$^{4}$ Joint Institute for Nuclear Astrophysics - Center for Evolution of the Elements, USA\\
$^{5}$ Department of Physics \& Astronomy, San Francisco State University, San Francisco CA 94132, USA \\
$^{6}$ Research School of Astronomy \& Astrophysics, Australian National University, Canberra, ACT 2611, Australia \\
$^{7}$ ARC Centre of Excellence for Astrophysics in Three Dimensions (ASTRO-3D), Canberra 2611, Australia \\
$^{8}$ INAF - Osservatorio di Astrofisica e Scienza dello Spazio di Bologna, Via Gobetti 93/3, I-40129 Bologna, Italy
}

\date{Accepted XXX. Received YYY; in original form ZZZ}

\pubyear{2015}

\begin{document}
\label{firstpage}
\pagerange{\pageref{firstpage}--\pageref{lastpage}}
\maketitle

\begin{abstract}
We combine stellar orbits with the abundances of the heavy, $r$-process element europium and the light, $\alpha$-element, silicon to separate in-situ and accreted populations in the Milky Way across all metallicities. At high orbital energy, the accretion-dominated halo shows elevated values of [Eu/Si], while at lower energies, where many of the stars were born in-situ, the levels of [Eu/Si] are lower. These systematically different levels of [Eu/Si] in the MW and the accreted halo imply that the scatter in [Eu/$\alpha$] within a single galaxy is smaller than previously thought. At the lowest metallicities, we find that both accreted and in-situ populations trend down in [Eu/Si], consistent with enrichment via neutron star mergers. Through compiling a large dataset of abundances for 54 globular clusters (GCs), we show that differences in [Eu/Si] extend to populations of in-situ/accreted GCs. We interpret this consistency as evidence that in $r$-process elements GCs trace the star formation history of their hosts, motivating their use as sub-Gyr timers of galactic evolution. Furthermore, fitting the trends in [Eu/Si] using a simple galactic chemical evolution model, we find that differences in [Eu/Si] between accreted and in-situ MW field stars cannot be explained through star formation efficiency alone. Finally, we show that the use of [Eu/Si] as a chemical tag between GCs and their host galaxies extends beyond the Local Group, to the halo of M31 - potentially offering the opportunity to do Galactic Archaeology in an external galaxy.

\end{abstract}




\section{Introduction}
\label{sec:intro}

The chemical fingerprint of a galaxy originates from elements forged in four main nucleosynthesis channels \citep{burbidge}. Very broadly, light, $\alpha$ elements are created in massive stars and dispersed by core-collapse (CC) supernovae (SNe), heavy elements up to iron and beyond are synthesised in both Type Ia SNe CCSNe and approximately half of the elements heavier than iron are made via the slow neutron capture process ($s$-process) in Asymptotic Giant Branch stars \citep{meyer94, busso99, kappeler11, karakas14}. The dominant site of the fourth channel, the rapid neutron capture process ($r$-process) responsible for the production of the other half of the elements heavier than iron, remains unconstrained \citep[see e.g.][]{Cowan1991, thielemann11, cote18, chiaki20}. 

Fortunately, Galactic Archaeology offers the chance to observationally constrain the yields of $r$-process elements in different environments potentially revealing the nature of this elusive channel. To achieve this goal, the population of low-metallicity, low-mass stars in the Milky Way must be unscrambled into distinct, co-evolved populations. Today, in the era of {\it Gaia} \citep{gaiadr2,gaiaedr3} and large-scale spectroscopic surveys, chemo-dynamical data is routinely used to pick out signatures of individual accretion events in the Galactic halo \citep[e.g.][]{belokurov2018, myeong2018s2, Helmi2018, Haywood2018, Koppelman2018,myeong2019, matsuno2019, yuan2020, monty2020, forbes2020, horta2021a, naidu2021, feuillet2021, malhan2022, buder2022, carrillo2022, ceccarelli2024}. 

Despite the high quality of the available data, neither dynamics nor chemistry alone is capable of unmixing the halo; both are needed in tandem to identify individual components with sufficient purity. This becomes particularly important when selecting members from the last major merger whose stars often overlap with in-situ populations in both orbital and chemical dimensions \citep[][]{Jean-Baptiste2017,pagnini2023}. While the view and origin of the Galactic halo remains complex and the literature is yet to agree on the number, timing, and mass of proposed accretion events, there is a general consensus that the last significant merger in the Galaxy's history was likely between the MW and the Gaia-Sausage/Enceladus (GSE) dwarf galaxy \citep[dGal, for a different view see,][]{donlon2020, donlon2022, donlon2023a}. 

Chemically, \citet{Nissen2010} presented one of the most striking lines of evidence for the existence of GSE (though it was not yet completely deciphered at the time) in the local view of the $\alpha$-metallicity plane. In their prolific study, stars assigned kinematically to the stellar halo are resolved into two distinct tracks, one with higher, and one with lower [$\alpha$/Fe] abundance ratio but overlapping broadly in [Fe/H]. It took the arrival of the {\it Gaia} data to reveal that the two halo sequences discovered by \citet{Nissen2010} also had different orbital and colour-magnitude properties \citep{Haywood2018, Helmi2018, Myeong2018action,Babusiaux2018}.

The distinct [$\alpha$/Fe]-abundance levels observed between GSE and the MW likely reflect differences in their star-formation histories. This interpretation stems from the assumption that $\alpha$ elements enter the interstellar medium (ISM) very quickly via CCSNe, after which they are recycled to form new stars, while the progenitors of SNIa, the main contributors of iron, need non-negligible time to form and merge. Type Ia SN explosions are therefore delayed with respect to both the star formation and CCSNe. Exploiting this delay, the abundance ratio of $\alpha$ to iron is used to indicate how fast the new stars are produced compared to the typical SN Ia delay time \citep{wallerstein1962, tinsley1979}. 

Simple assumptions about the recycling times for different element families can be combined with further approximations. Usually, consistency of the nucleosynthetic yield per explosion, the timing of mixing of the material injected into the ISM, as well as the gas circulation in and out of the galaxy, are all postulated to build powerful models of galactic chemical evolution \citep[][]{Weinberg2017,Andrews2017,Spitoni2017,omega}. Ultimately though, variations in abundance ratios generated by such models as a function of metallicity mostly reflect differences in otherwise constant element production yields modulated by galaxy's SF activity \citep[and/or the temperature of the ISM,][]{Schonrich2019}

Dynamically, GSE debris is associated with less bound energies, including some overlap with in-situ stars and has very little net spin, having lost most of its angular momentum in the interaction with the Galaxy \citep[see e.g.][]{belokurov2018,Helmi2018,naidu2021, vasiliev2022}. Many works have explored applying precise cuts in various parameter spaces to select pure samples of GSE and in-situ stars, relying primarily on the combination of the GSE's high-energy, plunging, radial orbits and its lower $\alpha$ abundance \citep[e.g. ][]{feuillet2021, buder2022}. In parallel to this pursuit, other studies have shown the risk (characterised as the contamination fraction) in adopting inconsistent selection criteria \citep{lane2022, carrillo2023}. Chemically, many efforts have been made to identify new ways in which GSE is unique relative to other MW stars at similar metallicities \citep{monty2020, matsuno2019, feuillet2021, buder2022, carrillo2022}, with light elements like aluminium (Al) emerging as a promising descriminator.

Concurrent to the discovery of GSE and other halo substructures, many studies have sought to associate halo Galactic Globular Clusters (GCs) with progenitor galaxies, assigning populations of GCs to individual accretion events \citep[e.g.][]{myeong2018a, myeong2019, massari2019, horta2020, forbes2020, callingham2022, limberg2022, monty2023b}. These efforts have thus far relied primarily on GC dynamics and/or their age-metallicity properties \citep{massari2023}, with limited success stemming from detailed GC chemistry. For example, a lack of clear differentiation in average light ($\alpha$) element abundances has been reported by \citet{horta2020}, while other studies demonstrate that going beyond a single $\alpha$ indicator, for example by adding [Al/Fe] information or heavy element abundances, may improve the quality of GC progenitor classification \citep[][]{limberg2022, monty2023b, belokurov2023a,belokurov2023b}. 

The main challenge associated with finding a chemical tag between GCs and their progenitors is the presence of anomalous chemical enrichment channels operating inside GCs \citep[e.g. see recent reviews by: ][]{bastian2018,Gratton2019,Miloni_Marino2022}. One example of this is the MgAl chain occuring during hot bottom burning, where Mg can be converted into Al, thus blurring or even destroying patterns otherwise established in the field stars \citep{karakas09}. Furthermore, it remains unclear whether GCs should share both chemical and dynamical coherence with their dGal progenitors \citep[e.g. they may be stripped earlier in the accretion process, while their hosts sink deeper into the MW potential, resulting in the two occuping different energies,][]{pagnini2023,chen2024}.

Recently, the $r$-process element europium (Eu) has emerged as a potential chemical tag to resolve differences between in-situ/accreted populations and importantly, it appears to be unaffected strongly by the chemical processes internal to GCs \citep{roederer11, mckenzie22, monty2023a, kirby2023}. In both GALAH survey data \citep{matsuno21, daSilva2023} and dedicated high resolution follow-up \citep{aguado21}, GSE stars have been found to contain an overabundance of Eu relative to inner halo MW stars of similar and higher metallicities. Because Eu is an almost pure $r$-process element \citep{bisterzo11}, the evolution of Eu as a function of metallicity in a range of environments can shed light on the physics of the $r$-process production and its deposition into the ISM. 

Importantly, the two main $r$-process production channels, CCSNe \citep[e.g. magnetro-rotational supernovae (MRSNe) and collapsars,][]{winteler12, tsujimoto15, siegel2019} and neutron stars (NS) as well as black hole neutron star mergers \citep[][]{Lattimer1974,rosswog14,Thielemann2017} have distinct delay times and ejecta energetics. NS merger delay times are not currently constrained and may be as large as 0.1-18 Gyr \citep{blanchard17, cote18, skulladotir19, skuladottir2020, naidu2022, reyes22, frebel2023}. Furthermore, predicted nucleosynthetic yields associated with the demographic of present-day binary neutron star systems in the MW may not be identical to the yields associated with populations which enriched at early times \citep{holmbeck2024}.

Studies extending $r$-process abundance analysis to GCs, which have been dynamically associated with GSE, find similarly high values of Eu \citep{kochhansen2021, monty2023a, monty2023b}. As mentioned, the prospect of chemically linking GCs to their progenitors using Eu relies on the Eu-enhancement in GCs being ``primordial'', and not the result of internal cluster evolution. This appears to be true in some clusters, where Eu-enhancement is independent of population within the cluster \citep[where populations are defined through their light element anti-correlations, see ][]{roederer11}.

For example, in $\omega$-Centauri ($\omega$-Cen), arguably the most complex `GC', \cite{johnson2010} find a plateau in [Eu/Fe] across all five populations (defined using their metallicity distribution functions.) Though this implies that self-enrichment of Eu in the cluster does occur, it goes with Fe-enrichment preserving the primoridal ratio found in the first generation of stars. In chemically complex GCs with much smaller Fe-spreads, a plateau in [Eu/Fe] is also found, e.g in M~22 \citep{marino11, mckenzie22} and NGC~362 \citep{monty2023a}. Note however that recent studies of M~92 \citep{kirby2023} and NGC~7078 \citep{Cabrera24} show evidence of correlations between Na and Eu in the first generation of stars. In the case of these clusters, Eu-enhancement is not interpreted as being primordial but is instead thought to be concurrent with GC formation.

In this study, we explore the use of Eu, weighted by a light $\alpha$-element, as a chemical tag differentiating \textit{both} in-situ and accreted field stars and globular clusters. Of the light $\alpha$-elements magnesium (Mg), silicon (Si) and calcium (Ca), we choose to neglect Mg in this study due to its strong involvement in GC light element anti-correlations.
Of the remaining light elements, Si and Ca, we choose Si as our light element tracer over Ca because it is a purer $\alpha$-element \citep{chiaki20}, giving access to three nucleosynthetic channels ($\alpha$, Fe-peak and $r$-process) across a broad range of metallicities (though with the caveat that it is involved in the MgAl chain in GCs, through Si-leakage.). After establishing the significance of this tag, we explore whether it extends across metallicities and investigate its origin using simple galactic chemical evolution models. Finally, we explore whether the tag extends beyond the MW halo into Local Group dwarf galaxies and their GCs and beyond, to the halo of M~31.

In Sec.~\ref{sec:dataset}, we present the sample of field stars we use, drawing from the third data release of the GALactic Archaeology with Hermes (GALAH) survey \citep{galahdr3}, the Measuring at Intermediate metallicity Neutron-Capture Elements
(MINCE) survey \citep{cescutti2022, francois2024} and the $R$-Process Alliance \citep{sakari2018, rpa4}. We also describe our compilation of literature GC abundances and new measurements in Sec.~\ref{sec:dataset}. In Sec.~\ref{sec:originofexcess} we explore the origin of the difference in [Eu/Si] between accreted and in-situ populations using a new one-zone chemical evolution model. We then discuss successes and failures of our model to fit the data and speculate as to the origins of the model-data disagreements. In Sec.~\ref{sec:globhist}, we show that GCs trace the star formation histories of their hosts across metallicities, in the MW halo, Local Group dwarf galaxies and potentially in the halo of M~31. Finally, we summarise our conclusions in Sec.~\ref{sec:conclusions}.

\begin{figure*}
	\centering
    \includegraphics[scale=0.7]{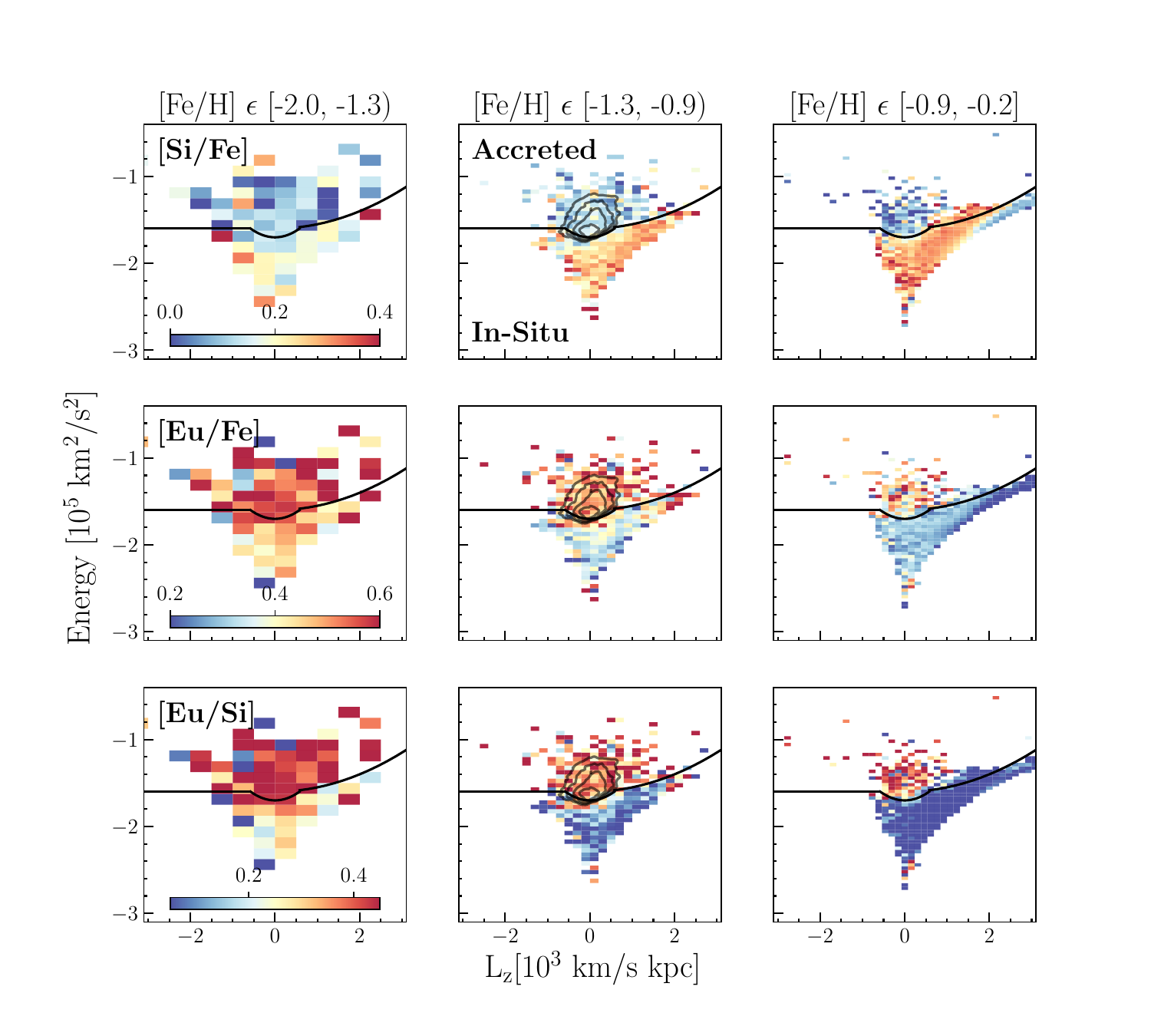}
	\caption{The distribution of our subset of GALAH DR3 stars (described in Sec.~\ref{sec:galah}) presented in Energy vs. $z$-component of angular momentum across metallicities. The metallicity range of GALAH stars is listed at the top of each column, while each row explores the evolution of [Si/Fe], [Eu/Fe] and the ratio of the two. The boundary in $E$-$L_{z}$ denoting accreted and in-situ stars is included and contours marking the extent of GSE are adopted from \citet{belokurovwrinkles} (and adapted to our choice of potential) and shown in the middle column. Note the increased density of Eu-enhanced stars in the region of E-L$_{\mathrm{z}}$ space occupied by GSE (emphasised in [Eu/Si] through an increased dynamical range.)}
	\label{fig:eumgelz}
\end{figure*}

\section{Chemical Abundance Data Set}
\label{sec:dataset}
In this section we discuss the field star data sets at across metallicities, including the selections we have made to acquire in-situ and accreted samples and discuss how the GC dataset was compiled. We present the abundance distributions for each sample in the corresponding section. 

\begin{figure*}
    \centering
	\includegraphics[width=\linewidth]{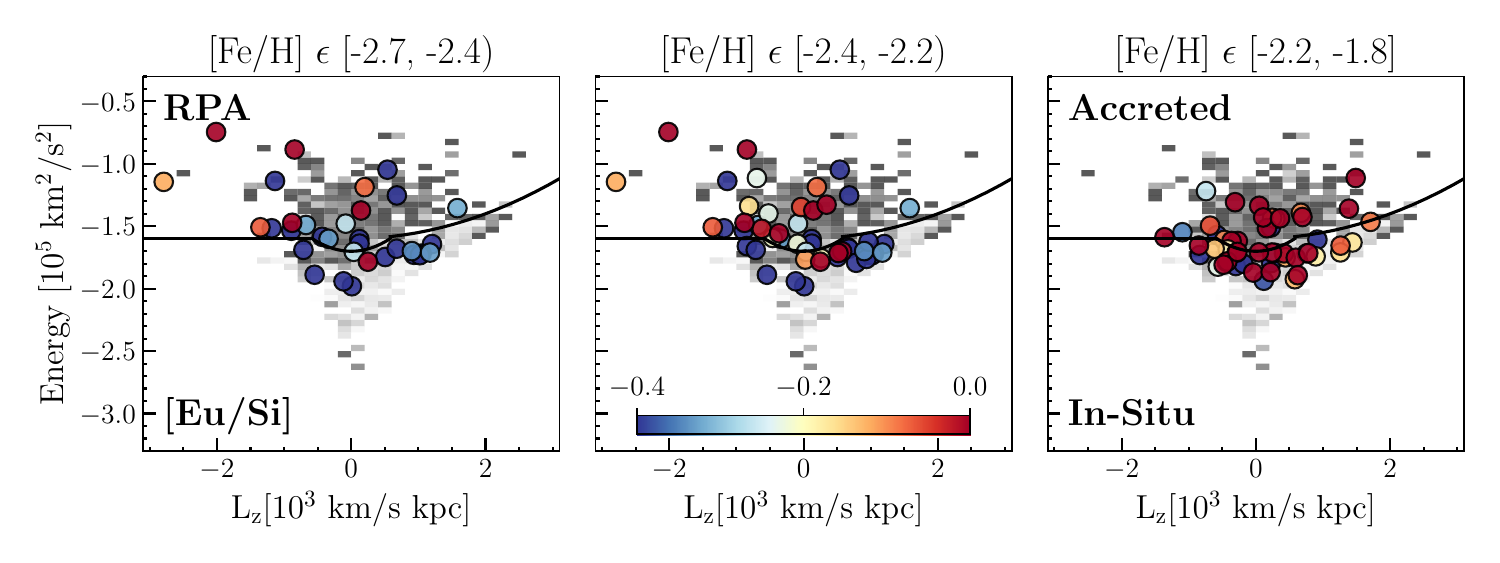}
    \caption{Stars from the combined, cleaned Northern Release of the R-Process Alliance DR1 data \citep{sakari2018} RPA DR4 catalogues \citep{rpa4} in $E$-$L_{z}$ space, coloured by their LTE ratio of [Eu/Si]. The underlying distribution of GALAH DR3 stars in the range of $\mathrm{-1.3\leq[Fe/H]}\leq-0.9$~dex is shown underneath. The combined sample is split into three metallicity bins around the mean NLTE-corrected metallicity ($\mathrm{[Fe/H]}=-2.2$). The left-most column only shows stars with a metallicity lower than the average minus 0.5$\sigma$.}
    \label{fig:eusielzrpa}
\end{figure*}

\subsection{\label{sec:fieldstars} Field Star Compilation}
\subsubsection{\label{sec:galah}GALAH Data}
To explore global trends in [Eu/Si] throughout the Galaxy, we utilise the GALAH DR3 data set of MW field stars \citep{galahdr3} above metallicities $\mathrm{[Fe/H]}\geq-2$. We choose this cut-off as the number of benchmark stars used to verify [Fe/H] in GALAH DR3 drops-off significantly below $\mathrm{[Fe/H]}=-2$. We follow the recommendations from GALAH regarding which flagged stars to remove\footnote{\url{https://www.galah-survey.org/dr3/using_the_data/}}. The following cuts are applied to only retain stars with an uncertainty in [EuSi/Fe] less than 0.2~dex, where \texttt{X} denotes either Eu or Si. In total we retain $\sim91,000$~stars from the original DR3 catalogue. It is important to note that non-local thermoydnamic equilibrium (NLTE) corrections have been applied to the published values of Si and Fe in GALAH DR3. Of the seven GALAH DR3 Si lines listed in \citet{amarsi2020} NLTE corrections, the average correction is $
\sim-0.04$~dex for metal-poor stars $-2\leq\mathrm{[Fe/H]}\leq-1$ across stellar types. NLTE corrections have \textit{not} been applied to Eu and all Eu abundances are derived from the the 6645~\AA\, line.

\begin{enumerate}
    \item \texttt{snr\_c3\_iraf}~$>30$
    \item \texttt{flag\_sp}~$=0$
    \item \texttt{flag\_fe\_h}~$=0$
    \item \texttt{e\_fe\_h}~$<=0.2$
    \item \texttt{flag\_X\_fe}~$=0$
    \item \texttt{e\_X\_fe}~$<=0.2$
\end{enumerate}
The dynamical properties, energy ($E$) and $z$-component of the angular momentum ($L_{z}$), are taken from the \texttt{GALAH\_DR3\_VAC\_dynamics\_v2} value added catalogue described in \citet{buder2022}. To calculate $E$, $L_{z}$ and other orbital parameters, \citet{buder2022} adopt the MW potential from \citet{mcmillan17}, assume a solar radius of 8.21~kpc and a circular velocity at the Sun of 233.1~km/s. They also orient the Sun 25~pc above the plane following \citet{juric2008} and adopt a total solar velocity of ($U, V, W$) = (11.1, 248.27, 7.25)~km/s in keeping with \citet{shonrich2010}. Throughout this study, we adopt this same reference frame and choice of the MW potential to derive dynamical properties for stars from other studies.

The [Si/Fe], [Eu/Fe] and [Eu/Si] distributions in the GALAH field star data set are presented in Fig.~\ref{fig:eumgelz}. The figure slices through $E$-$L_{z}$ space in bins of metallicity to present the evolution of Si, Eu and [Eu/Si]. The bin intervals are chosen such that the central bin captures the peak of the metallicity distribution of GSE \citep{Belokurov2022}. The size of the metallicity bin is listed at the top of the column in interval notation (where ``('' denotes an open interval). 

Two additional dynamical features are included in Fig.~\ref{fig:eumgelz} which we will revisit later. The first is the proposed boundary in $E$-$L_{z}$ between accreted and in-situ stars introduced by \citet{belokurov2023a}. This is marked with a thick black line and motivated in \citet{belokurov2023a} by light element variations seen in APOGEE \citep{apogeedr17} field stars. The second dynamical feature we include (in the middle column only where the bulk of GSE stars appear in GALAH) is a set of density contours which trace the extent of the region occupied by the debris associated the GSE merger \citep{belokurovwrinkles}. The exact energies for both dynamical features have been adjusted to accommodate our choice of potential. This was done by calculating the offset in energy between stars common to our study and that of \citet{belokurov2023a}.

Beginning in the highest metallicity interval (the right-most column in Fig.~\ref{fig:eumgelz}, a clear distinction is observed between the in-situ and accreted stars across both [Si/Fe] and [Eu/Fe]. The difference is likely driven by ``high metallicity'' stars from GSE ($\mathrm{[Fe/H]}>-0.9$), which dominates the accreted halo at these metallicities and the thick and thin discs (note the large number of stars at positive $L_{z}$, marking the prograde disc). Moving to intermediate metallicities (middle column), GSE more obviously dominates the halo, while the the in-situ component is now dominated by ``Splash'' stars \citep{belokurov2020} and the thick disc (some net rotation is still observed). Note again that the strong chemical differences persist between the two populations across all three element ratios.

Finally, in the lowest metallicity interval and left-most column, the range of Si and Eu abundances are similar across the in-situ/accreted boundary, but still display hints of possible distinction ($\Delta\mathrm{(above-below)}_{\mathrm{ave}}\sim0.2$~dex). This is likely because at low-metallicities, the signature of GSE becomes significantly weaker (exploring the low-metallicity tail of GSE), mixing with the signatures of smaller, earlier accretion events. In addition to this, the signature of the in-situ stars becomes less clear as we enter the population of ``Aurora'' stars with large chemical dispersion \citep{Belokurov2022}. However, despite this, combining the information provided by both Eu and Si into the ratio of [Eu/Si] leads to a slight increase in the abundance differences between accreted and in-situ ($\Delta\mathrm{(above-below)}_{\mathrm{ave}}\sim0.4$~dex).

\subsubsection{Low Metallicity Data Set}
\label{sec:lowmetal}
To explore the evolution of [Eu/Si] to metallicities below the range probed accurately by GALAH ($\mathrm{[Fe/H]}\leq-2$), we require additional data of low-metallicity field stars. To this end, we combined measurements of Eu and Si from the first Northern data release of the $R$-Process Alliance \citep[RPA DR2, ][]{sakari2018} and the fourth data release \citep[DR4,][]{rpa4}. We limit ourselves to only considering RPA survey stars when compiling our low-metallicity data set to minimise the introduction of additional zero point offsets. Unfortunately additional RPA releases do not include Si \citep{rpa1, rpa4holm}. We do not consider offsets between the GALAH and RPA data sets, however we never combine the two data sets directly (e.g. only the high-metallicity data set is used in the determination of a best-fit galactic chemical evolution model in Sec.~\ref{sec:gcemodels}.)

We apply a somewhat stricter cut on the quality of abundances to clean the combined RPA sample as Si lines become weaker and more difficult to measure at low-metallicities. We only retain stars with an uncertainty in [Eu/Si] less than or equal to the average uncertainty in [Eu/Si] ($\sigma\mathrm{[Eu/Si]_{ave}}\sim0.1$~dex, where $\sigma\mathrm{[Eu/Si]}$ is the quadrature sum of the uncertainties in Eu and Si.) Finally, given the stochastic nature of enrichment at low-metallicities, we only select stars with $\mathrm{[Ba/Eu]<0}$ to ensure Eu-enhancement traces the $r$ and not $s$-process channel. The final cleaned RPA sample contains 103 stars.

The published NLTE-corrected values of metallicity and $\mathrm{[X/Fe_{NLTE}]}$ are adopted for the RPA sample. NLTE corrections to [Eu/H] and [Si/H] have not applied by the RPA. Given the linelist of the RPA \citep{rpa4}, the NLTE corrections for Si are expected to be marginal \citep[$\sim-0.01-(-0.04)$~dex at $\mathrm{[Fe/H]}\sim-2$][]{amarsi2020}. Unfortunately, this is not necessarily the case for Eu. The two strongest Eu lines available in the optical occur at 4129~\AA\, and 6645~\AA. The bluer line is much more sensitive to NLTE corrections, with the average NLTE correction evolving nearly linearly from +0.1~dex at [Fe/H]$=0$, to +0.4~dex at [Fe/H]$=-4$ for a red giant branch star (T$_{\mathrm{eff}}=4500$~K, $\log~g=1.5$)\footnote{Nicholas Storm, private communication and Guo et al. in prep.}. The correction to the red line is less extreme, with corrections ranging from -0.05~dex at $\mathrm{[Fe/H]}=0$, down to 0.15~dex at $\mathrm{[Fe/H]}=-2$.

Eu-abundances in the RPA sample are predominantly determined from the stronger blue line of Eu at 4129~\AA\,, while the GALAH abundances are determined from the weaker Eu line at 6645~\AA\,. This difference necessary introduces some inconsistency. We discuss if this has any impact on our results in upcoming sections. Finally, to assign dynamical properties to the stars, we solve for the energies and $z$-angular momenta under the same MW potential \citet{mcmillan17} and LSR as was assumed for the GALAH dataset.

Fig.~\ref{fig:eusielzrpa} shows the distribution of the cleaned, combined RPA star sample in $E$-$L_{z}$ where each star is coloured by its corresponding the LTE [Eu/Si] abundance. The RPA data is plotted over-top of the intermediate metallicity GALAH sample, selected because it is better populated than the lowest metallicity bin. The combined sample is split into three metallicity bins to explore the evolution of [Eu/Si]. The central and right-most columns stars centred around the mean NLTE metallicity of the cleaned sample ($\mathrm{[Fe/H]}=-2.19$). The left-most column shows stars with a metallicity lower than the average metallicity minus 0.5 standard deviations. 

Three interesting observations emerge from Fig.~\ref{fig:eusielzrpa}. The first, is the overall tendency towards lower values of [Eu/Si] ($<0$~dex) with decreasing metallicity, common to both accreted and in-situ stars (note the range of the colourbar in Fig.~\ref{fig:eusielzrpa}). The second, is the appearance of a tentative split in the average [Eu/Si] value in the lowest metallicity bins across the accreted-in-situ boundary (labeled in the the third column of Fig.~\ref{fig:eusielzrpa}). On average, we find the accreted sample to be enhanced by $\sim0.13$~dex in [Eu/Si] relative to the in-situ sample. This is larger than the average measurement error in [Eu/Si] (neglecting upper limits in Eu).  We revisit the potential consequences of this potential split in accreted/in-situ populations in upcoming sections. 

The third interesting feature, is that despite the overall enhancement seen in [Eu/Si] in accreted stars relative to in-situ stars, not \textit{all} accreted RPA stars are enhanced in [Eu/Si] or [Eu/Fe] relative to their in-situ counterparts. This could be a reflection of the diversity in $r$-process enhancement observed in ultra-faint dGals (UFDs) \cite[e.g. as in the case of the UFDs Grus I and Triangulum II, in which neutron capture elements are not detected,][]{ji2019} and predicted through simulations of dGals \citep{kolberg2022}. It could also be a reflection of the transition in the dominance of the accreted halo away from GSE, towards smaller accretion events at lower metallicities. The appearance of [Eu/Si] enhanced in-situ stars could also reflect this second point, that earlier and/or less-massive accretion events could sink below the accreted/in-situ boundary marked by GSE - contaminating our in-situ selection. However, note that at low-metallicities (the central and left-most columns) the most [Eu/Si]-enhanced in-situ stars sit the closest to the boundary.

\subsubsection{Intermediate Metallicity Data Set}
After applying our cuts to both the GALAH and RPA data sets, we find a lack of stars in bins near the metallicity extremes of the two sets ($-2\leq\mathrm{[Fe/H]}\leq-1.6$. As such, we incorporate data from the Measuring at Intermediate metallicity Neutron-Capture Elements (MINCE) survey \citep{cescutti2022}, which specifically aims to fill this gap by measuring precise abundances in halo stars with $\mathrm{[Fe/H]}<-1.4$. We combine the first two publications from the MINCE group \citep{cescutti2022, francois2024}, apply the same cuts as GALAH (namely, uncertainty in [EuSi/Fe] less than 0.2~dex) and integrate the stars in our potential. In total, we retain 25 MINCE stars with orbital properties and [Eu/Si] abundances. To limit uncertainties associated with inconsistent Solar scales and abundance determinations, we restrict the overlap between the three samples when presenting the three together. Finally, we note that the MINCE collaboration measures Eu from both the 6645~\AA\, line (as in GALAH) and a line at 4435~\AA.

\subsection{Globular Cluster Compilation}

\subsubsection{Literature Data for 54 GCs}
\label{sec:litdat}
A selection of GC abundances of Eu and Si were compiled from the literature to sample both the accreted and in-situ populations defined in \citet{belokurov2023b}. We chose to prioritise GCs assigned to GSE in the literature, as GSE is likely the largest contributor of ex-situ GCs to the MW (especially at less-bound energies). Table~5 of \citet{johnson2017} was used as the starting point for the literature compilation. Emphasis was placed on building a compilation of studies with consistent wavelength coverage, instrumentation and spectroscopic analysis technique. As such, all the chemical abundances in the literature compilation are derived from optical data, in the wavelength range of $\sim3800-6700$\AA\, stemming predominately from VLT/FLAMES \citep{flames}, VLT/UVES \citep{uves}, Magellan/MIKE \citep{mike} or Keck/HIRES \citep{hires}. All abundance determinations were made under the assumption of Local Thermodynamic Equilibrium (LTE) and 35/51 studies explicitly state the use of MOOG \citep{moog} to perform either equivalent width analysis or spectrum synthesis. 

One of the major concerns regarding inconsistency across the compilation is the choice of line list, this includes the atomic data as well as the choice of isotopic splitting of the Eu line. To explore the severity of this, we synthesised Eu using the atomic data for the 6645~\AA\, line published in the oldest study in our set, \citep{mcwilliam1992}. Note that \citet{mcwilliam1992} was published before the study of \citet{lawler2001}, whose atomic data for Eu is commonly used in modern line lists \citep[e.g. when creating line lists via \texttt{linemake}][]{linemake}. While the excitation potential remained the same (1.38~eV), the log~$gf$ value differed by 0.08~dex between the two studies. Selecting the most Eu-rich star in NGC~2808 (to be discussed in the next section) as an example star, the change in log~$gf$ value resulted in a 0.04~dex change to the best-fit Eu abundance in the star, within measurement uncertainties. While this simple test is encouraging in the case of Eu, it does not address differences in the choice of isotopic splitting, which is often not explicitly stated in studies. We therefore cannot rule out inconsistent line lists as a source of uncertainty in creating our compilation.

Two other obvious inconsistencies in the compiled data set are the choice of Eu lines, and the solar scale. As discussed in Sec.~\ref{sec:lowmetal}, NLTE corrections to Eu vary significantly between the red, 6645~\AA\, and blue, 4129~\AA\, line. For simplicity and in-keeping with our field star compilation, we do not attempt to apply NLTE corrections to any of the literature data and therefore prioritise studies which use the red Eu line in their abundance determinations. Finally, although the studies have not been placed on the same solar scales, the photospheric abundances of Si and Eu only differ by $\pm0.04$ and $\pm0.01$ respectively between the popular solar scales of \citet{grevesse1998} and \citet{asplund09}. The LTE photospheric abundance of Fe is unchanged across the two studies. Therefore, we do not consider inconsistent solar scales to introduce significant uncertainties.

Our compiled dataset is presented in Table~\ref{tab:litcomp}. The adopted values of [Fe/H], [Si/Fe] and [Eu/Fe] and their respective uncertainties are listed alongside their designation of in-situ/accreted \cite[from ][]{belokurov2023b} and abundance references. In the case of multiple abundance uncertainties listed in the studies, we adopt the largest value listed (often the internal GC dispersion in each element). If more than one population of stars is present in the GC, we adopt the mean abundance for the most metal-poor population. If a metallicity spread is not obvious, we adopt the Mg-rich population. Both choices reflect a desire to sample the first generation of stars in the cluster. 

Measurements for NGC~2808 and NGC~1904 made in this study are included in Table~\ref{tab:litcomp} and discussed in Appendix~\ref{app:chemabundanceanalysis}. Note that this compilation was made prior to the publication of the study of $r$-process abundances in GCs made by \citet{schiappa2024}. We find good agreement with the Eu abundances published in their study for GCs in common, including between the two GCs analysed in this study, NGC~1904 and NGC~2808. We find good agreement within 0.06~dex in [Eu/Fe] for NGC~1904 between our two studies and acceptable agreement in NGC~2808 (0.10~dex difference). 

\begin{figure}
    \centering
	\includegraphics[width=\linewidth]{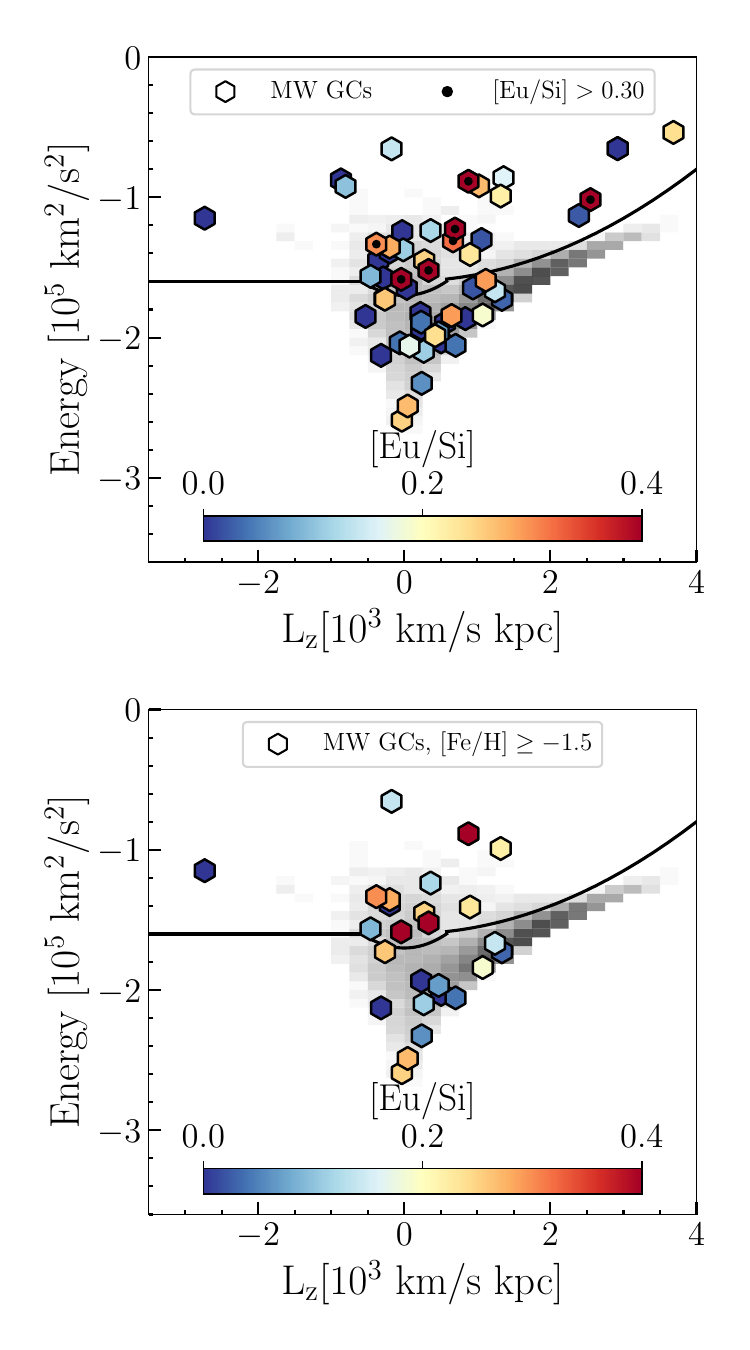}
    \caption{\textit{Top,} distribution of MW GCs (hexagons) in our sample over-plotted on top of the entire GALAH sample in log density, in $E$-$L_{z}$ space. The GCs are coloured by their [Eu/Si] abundance. GCs which exhibit an average [Eu/Si] abundance greater than 0.2~dex are marked with inset black circles. \textit{Bottom,} same as the above, but only GCs with metallicity greater than $\mathrm{[Fe/H]}\geq-1.5$ are included to highlight that the majority of [Eu/Si]-enhanced GCs occupy the region of $E$-$L_{z}$ space associated with accreted GCs.}
    \label{fig:eusielzgcs}
\end{figure}

\begin{table*}
    \centering
	\caption{\label{tab:litcomp} Literature compilation of GC chemical abundances used throughout this study. Note that all of the studies perform 1D, LTE analysis of individual stars using data from a high resolution spectrograph ($\mathrm{R}\sim40,000$). Assignment of in-situ/accreted (1/0) from \citet{belokurov2023b} is listed in the third-to-last column. Clusters with conflicting chemistry/dynamics are assigned ``0/1''. The literature source of each measurement is listed in the second-to-last column. }
    \begin{tabular}{lllllllcll}
        \hline
        GC & [Fe/H] & $\sigma$ & [Si/Fe] & $\sigma$ & [Eu/Fe] & $\sigma$ & In-Situ/Acc. & Study & Comment \\
           & & & & & & & (1/0) & &  \\
        \hline
        Arp	2       & -1.77 & 0.04 & 0.29 & 0.08 & 0.45 & 0.22 &  0  & \citet{mottini08} & ... \\
        NGC	104     & -0.67 & 0.05 & 0.30 & 0.10 & ... & ...   &  1  & \citet{carretta2004} & Si abundance only \\
        NGC	104     & ...      & ...     & ...     & ...     & 0.14 & 0.03 & ...  & \citet{james2004} & Eu abundance only \\
        NGC	1261    & -1.25 & 0.02 & 0.29 & 0.04 & 0.59 & 0.01 &  0  & \citet{kochhansen2021} & \\
        NGC	1851    & -1.18 & 0.07 & 0.39 & 0.03 & 0.67 & 0.11 &  0  & \citet{carretta1851} & \\
        NGC	1904    & -1.66 & 0.04 & 0.32 & 0.02 & 0.43 & 0.11 &  0  & This study & \\
        NGC	2298    & -1.91 & 0.10 & 0.51 & 0.05 & 0.34 & 0.15 &  0  & \citet{mcwilliam1992} & \\
        NGC	2419a   & -2.15 & 0.11 & 0.45 & 0.13 & 0.19 & 0.18 &  0  & \citet{cohen2012} & Mg-rich population\\
        NGC	2808    & -1.15 & 0.03 & 0.32 & 0.01 & 0.73 & 0.11 &  0 & This study & \\
        NGC	288     & -1.39 & 0.04 & 0.43 & 0.09 & 0.52 & 0.11 &  0/1  & \citet{ngc288ref} & \\
        NGC	3201    & -1.42 & 0.14 & 0.42 & 0.14 & 0.29 & 0.20 &  0  & \citet{ngc3201ref} & \\
        NGC	362     & -1.17 & 0.05 & 0.22 & 0.04 & 0.70 & 0.07 &  0  & \citet{carretta2013} & \\
        NGC 4147    & -1.86 & 0.01 & 0.46 & 0.01 & 0.37 & 0.03 &  0  & \citet{ngc4147ref} & \\
        NGC	4590    & -2.42 & 0.14 & 0.21 & 0.24 & 0.23 & 0.14 &  0  & \citet{ngc4590ref} & \\
        NGC	4833    & -2.25 & 0.09 & 0.74 & 0.09 & 0.36 & 0.13 &  1  & \citet{ngc4833ref} & \\
        NGC	5024    & -2.16 & 0.01 & 0.34 & 0.17 & 0.74 & 0.17 &  0  & \citet{lamb2015} & \\
        NGC	5139    & -1.77 & 0.11 & 0.29 & 0.16 & 0.23 & 0.22 &  0  & \citet{johnson2010} & Most metal-poor group\\
        NGC	5272    & -1.39 & 0.08 & 0.27 & 0.06 & 0.50 & 0.04 &  0  & \citet{ngc6205ref} & \\
        NGC	5286    & -1.80 & 0.05 & 0.40 & 0.01 & 0.32 & 0.03 &  0/1  & \citet{ngc5286ref} & \\
        NGC 5694    & -1.98 & 0.03 & 0.30 & 0.03 & 0.00	& ...  &  0  & \citet{ngc5694ref} & Eu upper limit \\
        NGC	5904    & -1.21 & ...  & 0.31 & 0.07 & 0.43 & 0.11 &  0  & \citet{ngc5904ref} & \\
        NGC 5927    & -0.47 & 0.02  & 0.24 & 0.03 & 0.43 & 0.04 &  1  & \citet{ngc5927ref} & \\
        NGC	6093    & -1.79 & 0.02 & 0.34 & 0.04 & 0.51 & 0.02 &  1  & \citet{ngc6093ref} & \\
        NGC	6121    & -1.18 & ...  & 0.57 & 0.08 & 0.34 & 0.10 &  1  & \citet{ngc6121ref} & \\
        NGC 6171    & -1.03 & 0.06  & 0.53  & 0.08  & ...   & ...   & 1  & \citet{carretta2009} & Si abundance only\\
        NGC 6171    & ... & ...  & ...  & ...  & 0.64   & 0.11   & ...  & \citet{schiappa2024} & Eu abundance only\\
        NGC	6205    & -1.50 & 0.07 & 0.31 & 0.11 & 0.57 & 0.11 &  1  & \citet{ngc6205ref} & \\
        NGC 6218    & -1.33 & 0.04  & 0.35  & 0.06  & ...   & ...   & 1  & \citet{carretta2009} & Si abundance only\\
        NGC 6218    & ... & ...  & ...  & ...  & 0.42   & 0.10   & ...  & \citet{schiappa2024} & Eu abundance only\\
        NGC 6254    & -1.58 & 0.06  & 0.28  & 0.05  & ...   & ...   & 1  & \citet{carretta2009} & Si abundance only\\
        NGC 6254    & ... & ...  & ...  & ...  & 0.52   & 0.10   & ...  & \citet{schiappa2024} & Eu abundance only\\
        NGC	6266    & -1.15 & 0.05 & 0.52 & 0.04 & 0.58 & 0.04 &  1  & \citet{ngc6266ref} & \\
        NGC	6273    & -1.77 & 0.08 & 0.35 & 0.10 & 0.39 & 0.15 &  1  & \citet{ngc6273ref} & \\
        NGC	6341    & -2.52 & 0.06 & 0.58 & 0.08 & 0.44 & 0.14 &  0  & \citet{kirby2023} & $\mathrm{[Mg/Fe]}>0.45$ \\
        NGC	6362    & -1.07 & 0.01 & 0.45 & 0.03 & 0.43 & 0.01 &  1  & \citet{ngc6362ref} & \\
        NGC	6388    & -0.37 & 0.09 & 0.32 & 0.10 & 0.21 & 0.08 &  1/0  & \citet{ngc6441ref} & Most-likely in-situ \\
        NGC 6388    & ...   & ...   & ...   & ...   & ...   & ...  & ... & ... & \citet{massari2023} and\\
        NGC 6388    & ...   & ...   & ...   & ...   & ...   & ...  & ... & ... & \citet{carretta2023}\\
        NGC 6397    & -1.99 & 0.04  & 0.34  & 0.05  & ...   & ...   & 1  & \citet{carretta2009} & Si abundance only\\
        NGC 6397    & ... & ...  & ...  & ...  & 0.63   & 0.10   & ...  & \citet{schiappa2024} & Eu abundance only\\
        NGC 6440    & -0.50 & 0.03  & 0.20  & 0.09  & 0.45 & 0.02 & 1 & \citet{ngc6440ref} \\
        NGC	6441    & -0.42 & 0.06 & 0.33 & 0.11 & 0.32 & 0.11 &  1  & \citet{ngc6441ref} & \\
        NGC 6528    & -0.20 & 0.06  & -0.02 & 0.05  & 0.25  & 0.03  & 1 & \citet{ngc6528ref} \\
        NGC 6553 & -0.14 & 0.06  & -0.06 & 0.09  & -0.02   & 0.04 & 1 & \citet{ngc6553ref} \\
        NGC 6584    & -1.52 & 0.06 & 0.33 & 0.06 & 0.65 & 0.11 &  0/1  & \citet{ngc7099ref} & \\
        NGC	6656    & -1.76 & 0.10 & 0.44 & 0.06 & 0.46 & 0.07 &  1  & \citet{marino11} & \\
        NGC	6715    & -1.55 & 0.03 & 0.11 & 0.07 & 0.38 & 0.17 &  0  & \citet{ngc6715ref} & \\
        NGC	6752    & -1.61 & 0.03 & 0.33 & 0.05 & 0.32 & 0.09 &  1  & \citet{ngc6752ref} & \\
        NGC	6809    & -2.01 & 0.02 & 0.50 & 0.01 & 0.54 & 0.02 &  1  & \citet{ngc6809ref} & \\
        NGC	6838    & -0.71 & 0.08 & 0.28 & 0.14 & 0.31 & 0.15 &  1  & \citet{ngc6838ref} & \\
        NGC	6864    & -1.16 & 0.08 & 0.37 & 0.09 & 0.62 & 0.14 &  0  & \citet{ngc6864ref} & \\
        NGC 6934    & -1.43 & 0.05 & 0.38 & 0.04 & 0.60 & 0.07 &  0  & \citet{ngc6934ref} & \\
        NGC 7006    & -1.55 & 0.03 & 0.26 & 0.05 & 0.36 & 0.05 &  0  & \citet{ngc7006ref} & \\
        NGC	7078    & -2.64 & 0.08 & 0.51 & 0.24 & 0.80 & 0.25 &  1/0  & \citet{ngc7078ref} & Large dispersion in [Eu/Fe]\\
        NGC	7078    & ... & ... & ... & ... & ... & ... &  ...  & ... & in \citet{Cabrera24}\\
        NGC	7089    & -1.68 & 0.04 & 0.40 & 0.01 & 0.38 & 0.04 &  0  & \citet{ngc7089ref} & Metal-poor group (``$r$-only'')\\
        NGC	7099    & -2.29 & 0.07 & 0.53 & 0.06 & 0.21 & 0.11 &  1  & \citet{ngc7099ref}& \\
        Pal	14      & -1.44 & 0.03 & 0.42 & 0.10 & 0.56 & 0.11 &  0  & \citet{pal14ref} & \\
        Pal	3       & -1.58 & 0.03 & 0.49 & 0.05 & 0.73 & 0.07 &  0  & \citet{pal3ref} & \\
        Pal	5       & -1.56 & 0.20 & 0.53 & 0.25 & 0.55 & 0.06 &  0  & \citet{pal5ref} & \\
        Terzan 7    & -0.61 & 0.04 & 0.00 & 0.01 & 0.53 & 0.05 &  0  & \citet{taut04} & \\
        \hline&     
	\end{tabular}
\end{table*}

\subsubsection{Archival Data for NGC~1904, NGC~2808 and NGC~1851}
While compiling the literature data set discussed in the previous section, we were unable at the time to find Eu abundance derivations for two potential GSE-GCs identified by both \citet{myeong2018a, myeong2019} and \citet{massari2019}, NGC~1904 and NGC~2808. To include these GCs in our compilation, we recovered high resolution archival VLT FLAMES/UVES \citep{flames, uves} spectra (program numbers 072.D-507 and 073.D-0211) for stars in each cluster and measured Eu, Mg, Si and Ca in both. The IDs of the stars are given in Table~\ref{tab:sps}, following the convention of \citet{carretta2009b}. The archival observations were previously used in the studies of \citet{carretta2009b} to extract light element abundances in a large number of MW GCs.

The spectra (for both the red and blue arms) were retrieved from the ESO Science Archive Facility. Four stars from each cluster with the highest average signal-to-noise ratios (SNR$\sim80$) were selected from each cluster. Prior to analysing the spectra, they were trimmed to remove discontinuities at the ends, continuum-fit with a low-order cubic spline using the \texttt{continuum} task in \textsc{IRAF}\footnote{IRAF is distributed by the National Optical Astronomy Observatory, which is operated by the Association of Universities for Research in Astronomy (AURA) under cooperative agreement with the National Science Foundation} and normalized before measuring the chemical abundances. The spectra were also radial velocity corrected using the \texttt{dopcor} task in \textsc{IRAF} and the radial velocities published in \citep{carretta2009b} prior to performing abundance analysis.

To extract chemical abundances, we performed standard 1D LTE equivalent width analysis and spectrum synthesis. Given the large catalogue of consistent GC abundances provided by \citet{carretta2004, carretta2007, carretta2009a, carretta2009b, carretta2013} in their series of prolific studies, and given that these catalogues dominate our literature sample, we examined the consistency of our chemical abundance analysis relative to the findings of \citet{carretta1851} and \citet{carretta2009b}. Furthermore, we performed the same analysis on the GC NGC~1851 for which \citet{carretta1851} also measure Eu, to examine the abundance consistency across elements. 

A description of the chemical abundance methodology and results of the consistency study are described in Appendix~\ref{app:chemabundanceanalysis}. In general, we find good agreement between our study and that of \citet{carretta2009b, carretta1851} across Si and Eu. In the case of Si, 9/12 stars display consistent (within measurement uncertainties) abundances across the three studies. In the case of our calibration cluster, NGC~1851, the values of Eu differ by on average only 0.09~dex between our study and that of \citet{carretta1851}.

\subsubsection{GCs enhanced in [Eu/Si] are also accreted}
Previously studies of the chemodynamics of low-metallicity field stars have revealed extreme $r$-process enhancement in a large number of accreted stars \citep{Sakari2018a, Roederer2018, Gudin2021, Shank2023, zhang2024}. Clustering of $r$-process enhanced stars purely in the halo purely in dynamics \citep[$E$-$L_{z}$ and eccentricity,][]{Gudin2021} has revealed groups with consistent chemistry \citep[small spreads in metallicity $\sigma<0.5$~dex,][]{Roederer2018} - suggesting a common origin. While some of these groups are associated with known GCs, others are not, pointing to the possibility for the existence of a number of disrupted low-mass dGals in the MW halo \citep[akin to the extremely $r$-process enhanced dGal, Reticulum II and explicitly seen in the $r$-process rich dGal stream, Indus,][]{Ji2016, Hansen2021}. While these studies have demonstrated a chemical distinctness in $r$-process elements in accreted field stars, systematic differences between large populations of in-situ and accreted GCs in $r$-process abundances has not been explored before.  

To investigate if a difference exists, we plot the GCs in our sample in Fig.~\ref{fig:eusielzgcs}, again in E-L$_{\mathrm{z}}$ space on top a 2D histogram showing the density of the entire GALAH sample. In the top panel we plot the entire sample of 54 GCs across metallicities, colouring each by the average [Eu/Si] abundance. GCs which exhibit an abundance of 0.3~dex or greater in [Eu/Si] are over-plotted using nested black dots. The dynamical separation between [Eu/Si]-normal and [Eu/Si]-enhanced GCs is striking. The the majority of the [Eu/Si]-enhanced GCs are accreted and show a slight bias to being prograde. Furthermore, note that a large number of [Eu/Si]-enhanced GCs occupy the region of E-L$_{\mathrm{z}}$ space associated with the GSE merger. This is highlighted more clearly in the bottom panel of Fig.~\ref{fig:eusielzgcs}, showing only the GCs with $\mathrm{[Fe/H]\geq-1.5}$.

\section{Origin of the [Eu/Si] Excess in the Accreted Halo}
\label{sec:originofexcess}
From Fig.~\ref{fig:eumgelz}, we have shown that the accreted halo appears significantly enhanced in [Eu/Si] relative to the remainder of the MW, arguably across all metallicities, but most obviously at intermediate metallicities. This is a known result \citep{matsuno21, aguado21} that we have presented in a different space through combining $E$-$L_{z}$ and abundances. When moving to the low-metallicity RPA sample, we find tentative evidence that the enhancement in [Eu/Si] found in accreted stars extends to NLTE metallicities as low as $\mathrm{[Fe/H]}\sim-2.5$ (LTE metallicities of $\mathrm{[Fe/H]}\sim-3$). To our knowledge, this is the first time this signature has been seen. Finally, we have shown that the most [Eu/Si]-enhanced GCs belong to the population of accreted clusters. In the upcoming section, we seek a potential explanation for this apparent enhancement in accreted stars using a galactic chemical evolution model. We discuss the strengths and weaknesses of our simplified model, speculating as to the cause of any disagreements and the potential temporal power encoded in the abundances of GCs.

\subsection{Selecting In-Situ and Accreted Field Stars}
To consider the evolution of accreted and in-situ field stars across metallicity, we split our GALAH, MINCE and RPA field stars into accreted and in-situ populations. The mean values of these two groups will then be used to constrain our GCE models. Beginning with the GALAH data, we select accreted stars by sampling the centroid of the GSE contours in $E$-$L_{z}$ space presented in the middle column of Fig~\ref{fig:eumgelz}. We do this by drawing a circle radius radius 0.2 in $E/10^{5}$, $L_{z}/10^{3}$ and selecting stars in the metallicity range $-2\leq\mathrm{[Fe/H]}<-0.6$. From this sample, further chemical cuts are made such that bonafide GSE stars must follow $\mathrm{[Mg/Fe]}\leq-0.3\times\mathrm{[Fe/H]}$ \citep{belokurov2023a, belokurov2023b}. Our final GSE GALAH sample contains $\sim860$ stars.

To select in-situ stars in our GALAH sample, we again select stars in $E$-$L_{z}$, but draw from deeper in the MW potential. We select stars within a circle radius 0.2 in $E/10^{5}$, $L_{z}/10^{3}$ centred at [0, -2.25] and require only that $\mathrm{[Fe/H]}\geq-2$. The in-situ GALAH sample selected contains $\sim5300$ stars. To extend to lower metallicities, we also classify the MINCE and RPA samples into accreted and in-situ. To do this, we simply divide the samples into ``accreted'' and ``in-situ'' using the boundary marked in Fig~\ref{fig:eusielzrpa} and described in \citet{belokurov2023a}. Stars above the boundary are deemed accreted, and stars below, in-situ. Of the 25 MINCE stars in our sample, we classify 15 as accreted and ten as in-situ. Of the 99 RPA stars in our sample, 46 are classified in-situ and 53 are classified as accreted. We note that the sample of in-situ MINCE and RPA stars is likely contaminated by the most-bound accreted stars (likely accreted at very early times). However, until a more complete low-metallicity sample of in-situ halo stars exists, we are forced to adopt this simple cut. 

We present our sample of accreted and in-situ field stars in Fig.~\ref{fig:gcemodels} in red and blue respectively. This colour convention is maintained throughout. We plot the binned mean in Fig.~\ref{fig:gcemodels}, using bins of size 0.3~dex for the GALAH samples, 0.9~dex for the MINCE samples and 0.4~dex for the RPA samples. The standard deviation of each sample is shown as the large shaded region, while the median absolute deviation weighted by the number of stars per bin is shown as the smaller, darker shaded region. The GALAH samples maintain separation across the three element ratios ([Si/Fe], [Eu/Fe] and [Eu/Si]). Note that at metallicities lower than [Fe/H]$\sim-1.3$~dex in the in-situ selection, we are likely selecting ``Aurora'' stars, which is speculated to be the primordial component of the MW \citep{Belokurov2022,myeong2022}. The sustained difference between Aurora and GSE in [Eu/Si] supports different star formation histories for the two ancient galaxies. The low-metallicity combined MINCE and RPA samples also show a separation between in-situ/accreted across the three element ratios. The separation of the two in [Eu/Si] as hinted at in Fig.~\ref{fig:eusielzrpa} appears once more. We note the combination of Si and Eu creates a larger dynamical range over which the two samples separate - highlighting its diagnostic power.

\subsection{Galactic Chemical Evolution Models}
\label{sec:gcemodels}

To interpret the presented data samples, we construct simple analytic chemical evolution models as discussed by \citet{Weinberg2017} and \citet[Sanders, in prep.]{Sanders2021}. For this modelling, we assume that (i) the star formation history (SFH) is linear-exponential ($\dot{M}_\star\propto t\,\mathrm{exp}(-t/\tau_\mathrm{sfh})$), (ii) the star formation efficiency (SFE) is constant $1/\tau_\star=\dot{M}_\star/M$, (iii) the mass-loading factor is constant leading to constant depletion time, $\tau_\mathrm{dep}$, (iv) the stellar products enter a single cold interstellar medium (ISM) phase after some delay time distribution depending on the specific channel and (v) the stellar yields from each channel are metallicity independent. 

We consider three stellar yield channels: (i) core collapse supernovae (CC) that return Mg, Si and Fe to the ISM instantly, (ii) type Ia supernovae (Ia) that return Si and Fe to the ISM after a delay-time distribution $\propto t^{-q_\mathrm{Ia}}$ for $t>t_\mathrm{D,Ia}$ and (iii) NS mergers that return Eu to the ISM after a delay-time distribution $\propto t^{-q_\mathrm{NS}}$ for $t>t_\mathrm{D,NS}$. These delay-time distributions are handled in an approximate analytic way using a sum of three exponential functions \citep[each of which is analytically tractable,][]{Weinberg2017}. The production of element $x$ from each stellar yield channel $\mathrm{Y}$ is parametrized by $m_x^\mathrm{Y}$, the mass of the element returned after an infinite time per unit mass of stars formed. We work with these quantities on the `solar scale' (as reported in the data) i.e. normalized by the mass fraction of element $x$ in the Sun as reported in \citet{asplund09}. 

Note that we do not consider MRSNe as a source of $r$-process in our models, removing the prompt enrichment channel. This choice is largely driven by the rise in [Eu/Si] seen in our compilation at low-metallicities which is in contrast to the plateau predicted by MRSNe. We explore including MRSNe in our models in an upcoming companion study (Sanders et al. in prep).

\begin{table}
    \caption{Adopted priors for the chemical evolution modelling. Each parameter has a normal prior with the given means and standard deviations (S.D.). The top section contains system-dependent parameters (indexed by $s$) and the lower section system-independent parameters.}
    \centering
    \begin{tabular}{lcc}
    \hline
    Parameter&Mean&S.D.\\
    \hline
    $\log_{10}\frac{\displaystyle m_{\mathrm{Fe}}^\mathrm{CC}}{\displaystyle \tau_{\star,s}[\mathrm{Gyr}]}$&$-3.45$&$0.27$\\
$\tau_{\mathrm{sfh},s}$&$10\,\mathrm{Gyr}$&$10\,\mathrm{Gyr}$\\
$\tau_{\mathrm{dep},s}$&$2\,\mathrm{Gyr}$&$2\,\mathrm{Gyr}$\\
$t_{\mathrm{max},s}$&$10\,\mathrm{Gyr}$&$10\,\mathrm{Gyr}$\\
$\ln\sigma_{a,s}$&$-3$&$1$\\
\\
$\log_{10} m_x^\mathrm{Y}$&$0$&$\infty$\\
$[\mathrm{Mg}/\mathrm{Fe}]^\mathrm{CC}$&$0.4$&$0.05$\\
$\log_{10}(m_\mathrm{Fe}^\mathrm{Ia}/m_\mathrm{Fe}^\mathrm{CC})$&$0.32$&$0.23$\\
$t_\mathrm{D,Ia}$&$60\,\mathrm{Myr}$&$20\,\mathrm{Myr}$\\
$t_\mathrm{D,NS}$&$10\,\mathrm{Myr}$&$3\,\mathrm{Myr}$\\
$q^\mathrm{Ia/NS}$&$1.1$&$0.2$\\
\hline
    \end{tabular}
    \label{tab:priors}
\end{table}

We jointly model both the accreted and in-situ GALAH datasets at high metallicities \textit{only}, due to the uncertainty in GSE membership at low-metallicities. As further clarification, because we are jointly modeling the two data sets, the chemical yields are fixed and the models may only alter the SFHs and depletion timescales (outflows) of the two systems to fit the data. For each dataset, we bin the stars into $10$ equally-populated bins in $[\mathrm{Fe}/\mathrm{H}]$ and compute median abundances for the $i$th bin $\mathcal{A}^i=([\mathrm{Fe}/\mathrm{H}]^i,[\mathrm{Mg}/\mathrm{Fe}]^i,[\mathrm{Si}/\mathrm{Fe}]^i,[\mathrm{Eu}/\mathrm{Fe}]^i)$ and corresponding uncertainties $\sigma^i_\mathcal{A}$ as $1.4826\times\mathrm{MAD}$ (approximately one standard deviation) where $\mathrm{MAD}$ is the median absolute deviation. For each set of chemical evolution model parameters, we compute the evolution of abundance $a$ as $\tilde{\mathcal{A}}_a(t)$ and calculate the product of individual datum likelihoods 
\begin{equation}
\mathcal{L}^i=t_\mathrm{max}^{-1}\int_0^{t_\mathrm{max}}\mathrm{d}t\,\prod_a \mathcal{N}(\mathcal{A}_a^i|\tilde{\mathcal{A}}_{a}(t),{\sigma^{i}_{\mathcal{A}, a}}^2+\sigma^2_{a}),   
\end{equation}
where product $a$ is over abundances. $\sigma_{a}$ are introduced to account for additional scatter and we set $\sigma_{[\mathrm{Fe}/\mathrm{H}]}=0$. Each datum has an unknown time $t$ that we marginalize over. Note we do not weight the data by the star formation history to mitigate the impact of selection effects (and our binning) on the modelling.

The models are implemented in the probabilistic programming language, \textsc{Stan} \citep{stan}. We fit for a set of system-dependent parameters for the $s$th system, $(m_{\mathrm{Fe}}^\mathrm{CC}/\tau_\star, \tau_\mathrm{sfh}, \tau_\mathrm{dep}, t_\mathrm{max})_s$, and the `universal' stellar yields from each channel, $m_x^\mathrm{Y}$, which are system-independent. We use the priors listed in Table~\ref{tab:priors}. Note that $m_{\mathrm{Fe}}^\mathrm{CC}/\tau_\star$ is chosen as the parametrization of the star formation efficiency because it is this quantity that governs the metallicity scale of the models and without prior information, element production and star formation efficiency are degenerate. The prior on $\ln m_\mathrm{Fe}^\mathrm{CC}/\tau_\star$ has been chosen using core-collapse supernova yields tables \citep{WoosleyWeaver1995, ChieffiLimongi2004, Kobayashi2006, Nugrid1, Nugrid2} and reasonable initial mass functions \citep[e.g.][]{Kroupa2001, chabrier}, to find $\log_{10}m_\mathrm{Fe}^\mathrm{CC}=(-3.17\pm 0.17)$ combined with the star formation efficiency measurement of $1/\tau_\star=(5.25\pm 2.5)\times10^{-10}\,\mathrm{yr}^{-1}$ from \cite{Leroy2008} (which we convert to the solar abundance scale using the solar mass fraction $\log_{10}Z_{\sun,\mathrm{Fe}}=-2.9$). 
We place a prior on the relative iron production from Type Ia and CC SNe using the number of Type Ia produced per unit stellar mass formed of $N_\mathrm{Ia}/M_\star=(2\pm0.5)\times10^{-3}$~M$_\odot^{-1}$ from \cite{MaozMannucci2012} and an approximate $M_\mathrm{Fe}=(0.7\pm0.15)$~M$_\odot$ of iron produced per Type Ia. We further place a prior on $[\mathrm{Mg}/\mathrm{Fe}]^\mathrm{CC}$ to encourage the models to reproduce reasonable early plateaus in $[\alpha/\mathrm{Fe}]$.

\subsection{Star Formation efficiency alone cannot explain the difference between GSE and the MW}
\begin{figure}
	\centering
    \includegraphics[width=\linewidth]{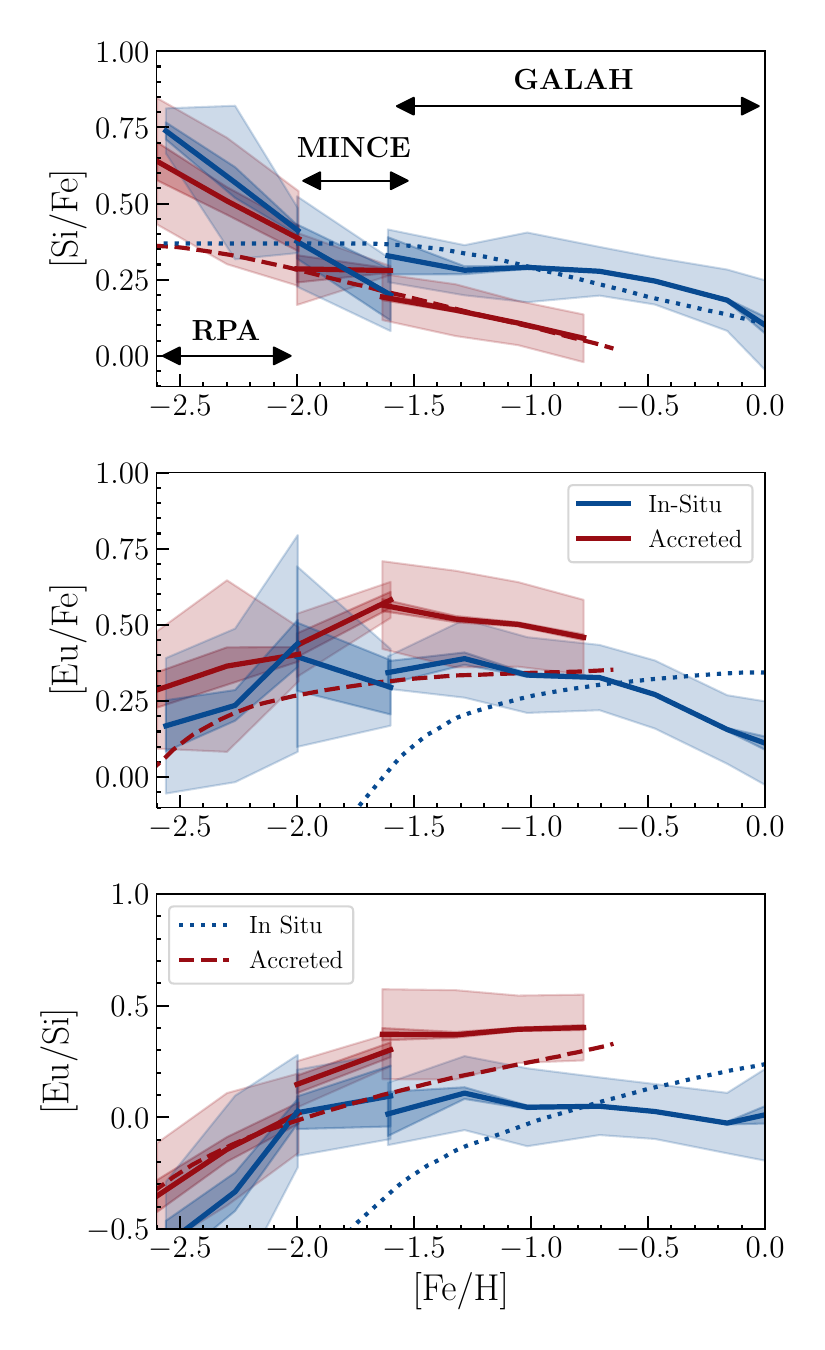}
	\caption{Predictions from our galactic chemical evolution models for the in-situ (dotted line) and accreted (dashed line) jointly fit to the GALAH populations. The predictions are shown across the three ratios, alongside the running mean, the median absolute deviation weighted by the square root of the number of stars per bin (dark shaded region) and the standard deviation of the GALAH, MINCE and RPA samples (shaded region). The accreted data is shown in red and the in-situ in blue.}
	\label{fig:gcemodels}
\end{figure}

The GCE models described in the previous section are presented in Fig.~\ref{fig:gcemodels} across the three element ratios. Recall that the models were jointly fit, and only fit to the in-situ and accreted GALAH field star data (above $\mathrm{[Fe/H]}\geq-2$) in [Si/Fe], [Mg/Fe] and [Eu/Fe]. The best-fit accreted GCE model is marked with dashed blue line in all panels, while the best-fit in-situ model is marked with a red dotted line.

While the models do a great job at replicating the two populations in [Si/Fe] at high metallicities ($\mathrm{[Fe/H]}>-2$), neither model does a good job of fitting its respective component at low-metallicities. This is perhaps most perplexing in the ratio of [Si/Fe], where the RPA data for both components is significantly higher than predictions. Examining the [Mg/Fe] abundance ratios presented in fig.~10 of \citet{sakari2018}, the [Si/Fe] and [Mg/Fe] abundances seem to be generally consistent. However, at $\mathrm{[Fe/H]}\sim-3$, the mean [Mg/Fe] abundance is $\sim0.5$~dex, 0.25~dex lower than the average value of [Si/Fe]. A rise is $\alpha$-elements at low-metallicities has been seen in dedicated high resolution studies of [O/Fe] \citep[e.g. in both 1D LTE and 3D NLTE abundances][]{amarsi2019}, but is not recovered by large surveys like APOGEE \citep{johnson2020}. Regardless of the fit to the low-metallicity [Si/Fe] values, the high metallicity end shows a good fit - helping to pin down the star formation efficiencies of the two systems by way of the $\alpha$-knee.

Moving to [Eu/Fe] and [Eu/Si], the most revealing feature of the models is not their inability to fit the individual populations, but rather their inability to replicate the trends seen in both. Recall that because the models are jointly fit to both populations, the only free parameters which may be adjusted to fit the data are i) the SFH of each progenitor and, ii) outflows from the system. The metallicity scale predicted by the models and governed by the SFE determined in the fit to [Si/Fe], predicts the appearance of a peak in [Eu/Fe] at higher metallicities than what is seen in the data. This is the case for \textit{both} populations. Furthermore, while the models predict the two populations will reach the same plateau in [Eu/Fe], this is clearly not the case in the data. Finally, while the models predict a continued climb in the ratio of [Eu/Si], both data sets display a plateau across metallicities from $-1.5\leq\mathrm{[Fe/H]}\leq0$.

Considering both the successes and failures of our jointly fit models across the two populations, we interpret, i) the high-quality fit to [Si/Fe] informing us that SFE alone cannot explain the differences between the two populations, as the metallicity scale set by this SFE fails to predict the evolution of [Eu/FeSi] and, ii) the appearance of the different plateaus in the two populations is the result of the two galaxies reaching different equilibrium levels of [Eu/Si]. Finally, we interpret the upwards trend of [Eu/Si] at low-metallicities as support for NS mergers dominating $r$-process enhancement at early times. This last point is explored further in an upcoming study (Sanders et al. in prep).

In both the studies of \citet{matsuno21} and \citet{naidu2021}, the evolution of [Eu/Mg] is compared across MW stars with different origins. In \citet{matsuno21}, GSE stars are compared against in-situ MW stars, while \citet{naidu2021} compare GSE stars to stars from the proposed \textit{Kraken} merger, theorised to have taken place early on in the MW's history \citep{kraken}. Regardless of the true nature of the non-GSE component, given the selection space of the in-situ and \textit{Kraken} stars in both \citet{matsuno21} and \citet{naidu2021}, overlap in samples is likely. While both studies attribute the difference between populations to the different star formation histories, their explanations differ slightly. \citet{matsuno21} argue a more efficient and extended star formation history is needed to explain the low [Eu/Mg] values in the MW at higher metallicities, while \citet{naidu2021} argue a shorter star formation duration explains the lower value in \textit{Kraken} (which was truncated before achieving the value seen in GSE).

While the explanation of \citet{naidu2021} seems plausible if both selections originate in dGals, it is incompatible with the explanation of \citet{matsuno21} and with data found in other dGals at similar metallicities. \citep[e.g. in the LMC where $\mathrm{[Eu/Mg]}\sim0.4$~dex at $\mathrm{[Fe/H]}\sim-1.5$ despite large differences in the predicted SFH and SFE between the LMC and GSE,][]{vanderbarlmc, reggiani2021}. If we accept the explanation of \citet{matsuno21}, that star formation efficiency alone explains the offset (merely a shift in the x-axis due to the MW producing stars to higher metallicities), we would expect the in-situ MW stars to trend up, increasing in [Eu/Si] at higher metallicities. In fact, this is the trend predicted by our simple GCE models. In an upcoming paper (Sanders et al., in prep), we investigate whether multi-zone ISM models can reproduce the flat trends observed in the data.

\subsection{Implications of the potential continuation of differences between in-situ and accreted stars below $\mathrm{[Fe/H]}\sim-2$}
To explore the robustness of a potential continuation in the difference of [Eu/Si] between accreted and in-situ samples at low-metallicities, we perform a A Kolmogorov-Smirnov (KS) test using \textsc{Scipy} \texttt{stats.ks\_2samp} on the two populations in the combined RPA sample. The test returns a global \textit{p}-value of 0.07 when considering the distributions of [Eu/Si], or a $\sim1.8\sigma$ confident rejection of the hypothesis that the two are drawn from the same distribution. If we perform a sliding window KS-test as described in \citet{davies2024} using a window size of 0.4~dex, we find $p$-values less than or equal to 0.05 ($2\sigma$ confidence) for the two populations below $\mathrm{[Fe/H]}=-2.45$. Finally, if we cull the RPA sample to keep only giant stars with $\log~g<3$ and $\mathrm{T_{eff}}\leq5500$ to apply the linear NLTE correction to Eu mentioned in Sec.~\ref{sec:lowmetal}, and perform the same test, we find a global \textit{p}-value of 0.04, or $>2\sigma$ confidence that two samples did not originate in the same distribution. 

This is in contrast to what has been seen in the literature before, both through simulations and observations. In \citet{skuladottir2020}, a flat trend in [Eu/Mg] is observed across metallicities in MW field stars. At metallicities of $\mathrm{[Fe/H]}<-1$, \citet{skuladottir2020} adopt the abundances of \citet{roederer2014a} which, like the RPA data, sit both above and below the accreted-in-insitu boundary in $E$-$L_{z}$ (when placed in our same potential). Examining the [Eu/Mg] distribution of the two populations in \citet{roederer2014a} (limited by the small number of stars with well-measured [Mg/Fe] and Gaia data, $n=72$) we again find that the accreted sample is enhanced in [Eu/Mg]. Performing the same KS-Test (as-in the RPA sample) we find a \textit{p}-value of $4.6\times10^{-5}$ ($>99\%$ confidence that the two samples are not drawn from the same distribution.)

Across the metallicity interval $-2.2\leq\mathrm{[Fe/H]}\leq-1$ where \citet{skuladottir2020} utilise the \citet{roederer2014a} data, the in-situ sample has an average value of $\mathrm{[Eu/Mg]}\sim-0.3$, in contrast to the accreted sample at $\mathrm{[Eu/Mg]}\sim0$. This is in agreement with our finding, that the ratio of [Eu/$\alpha$] in the in-situ population decreases at higher metallicities more rapidly than the accreted population. In summary, if the two populations are not resolved, this would lead to the interpretation that the MW (in-situ) population maintains a flat mean abundance in [Eu/Mg] ($\sim0$) with large scatter.

Some analytical galactic chemical evolution models have assessed their validity in reproducing the MW by capturing the apparently flat trend of [Eu/$\alpha$] or [Eu/Fe] with time. The model of \citet{kobayashi2020}, a one-zone model with outflows and many production sites, predicts the flat trend of [Eu/Fe] and [Eu/O] at low-metallicities \citep{kobayashi2023}. When comparing the model to observational datasets of halo stars, the flat trend fits the data well because of the apparently large dispersion at low-metallicities. 

In addition to using the flat trend to constrain the global SFH of the MW, the analytical model of \citet{kobayashi2023} also uses the apparent trend to determine the rate of magneto-rotational SNe in their models. Even if we apply the linear NLTE correction discussed in Sec.~\ref{sec:lowmetal} associated with the blue line of Eu, we find that the ratio of [Eu/Si] continues to decrease in both populations at low-metallicities. While magneto-rotational SNe enrich in both $\alpha$ and $r$-process producing the flat trend in [Eu/$\alpha$], our result implies that a more pure $r$-process channel like NS mergers better predicts the evolving trend seen in [Eu/Si].

The origin of the large scatter in [Eu/Fe] and [Eu/$\alpha$] at low-metallicities in the MW has recently been probed using cosmological simulations \citep{haynes2019, vandevoort2020, hirai2022}. In \citet{vandevoort2020}, they introduce $r$-process elements into 16 MW-like galaxies from the AURIGA simulation suite \citep{auriga1}. They do this by implementing the $r$-process elements as ``passive tracers'', meaning that they are released into the interstellar medium (ISM) without impacting the dynamical evolution of the galaxy. They implement two sites, rare core collapse SNe and neutron star mergers stochastically, fixing the yields from each across all metallicities.

To assess the agreement between their model and observations of MW stars, \citet{vandevoort2020} look at the evolution of the mean abundance of [Eu/Fe] for both disc and halo stars (at $z=0$) in their simulations. To compare to observations they use low-metallicities stars from the SAGA database \citep{saga}, undoubtedly composed of both in-situ and accreted stars. However, \citet{vandevoort2020} know the origin of their MW stars exactly, finding only 8\% of the total stars in their selection to have an ex-situ origin. This number increases substantially at low-metallicities, with 78\% of the stars below $\mathrm{[Fe/H]}<-2$ having formed ex-situ - the metallicity at which the dispersion in [Eu/Fe] increases. Resolving abundance differences in low-metallicity observational datasets between accreted/in-situ stars would make it possible to exploit knowledge of the origin and thus, the star formation history of the progenitors provided by simulations.

Ultimately, \citet{vandevoort2020} are unable to determine whether their fiducial NS merger model, which predicts a rise in [Eu/Fe] with metallicity at early times, and their rare CCSNe model, which predicts a flat value of [Eu/Fe] across metallicities, better reflects the observations. This is also the case in \citet{haynes2019}. While the presence of large dispersion in [Eu/Fe] and [Eu/$\alpha$] at low-metallicities can be reproduced in hydrodynamical simulations, decoupling the physical source of the dispersion (site and efficiency of $r$-process production) from the accretion history of the galaxy may not be possible without knowing the origin of the stars (and the unique star formation histories of their progenitors). That is, without knowing the fractional contribution of the in-situ and accreted components in observational data, it is likely impossible to disentangle the physical source of the dispersion from the selection function of the study.

Recently, \citet{ou2024} also observed a downwards trend in [Eu/Mg] in GSE stars at low-metallicities, however they have relatively few (eleven) stars in their study below $\mathrm{[Fe/H]}=-2$ (it is unclear whether this is an LTE or NLTE metallicity). Below $\mathrm{[Fe/H]}=-2$ they find a large dispersion in [Eu/Mg], which they interpret as being due to delayed $r$-process enrichment with plausible delay times ranging from 10 - 500~Myr. While we likely need more bonafide in-situ stars at low-metallicities to explore the significance of the difference between populations, we are confident in the general downwards trend we also find in both populations at low-metallicities. As discussed in the next section, accreted GCs in our dataset, specifically those identified by 
\citet{myeong2019} as belonging to GSE, trace the same trend seen in the field stars - namely a downwards trends with decreasing metallicities, seen in both [Eu/Si] and [Eu/Fe]. Furthermore, the GCs provide tight abundance constraints in both [Eu/$\alpha$] and [Fe/H]. Finally, the agreement seen in the GCs is not the result of Eu, nor Si, in isolation, but the evolution of both. We discuss possible extensions of this discovery in the next section. 

\begin{figure*}
	\centering
    \includegraphics[width=\linewidth]{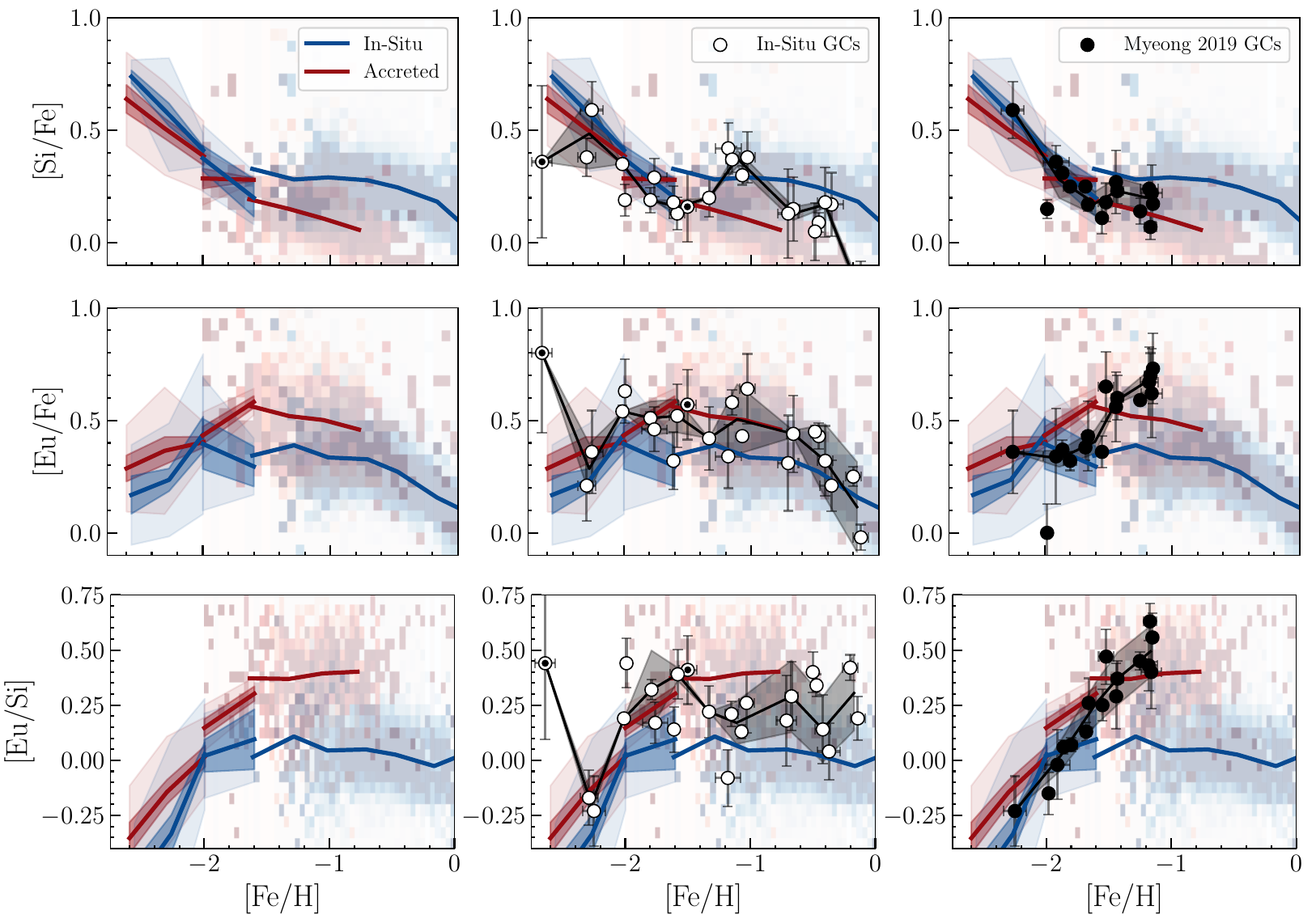}
	\caption{\textit{All,} abundance of [Si/Fe], [Eu/Fe] and [Eu/Si] for the in-situ and accreted samples of field stars. The column-normalised in-situ and accreted samples of GALAH stars are shown as blue and red clouds respectively, with the binned mean value over-plotted. The intermediate metallicity MINCE and low-metallicity RPA data sets are also included, with the red being stars found above the \citet{belokurov2023a} E-L$_{\mathrm{z}}$ cutoff and the blue below. In the case of the RPA data and MINCE data, the binned mean value is plotted along with the standard deviation weighted by the square root of the number of stars per bin (dark shaded region) the standard deviation in each bin (lighter shaded region). \textit{Centre,} in-situ GCs as classified by \citet{belokurov2023b} are plotted over-top of the MW populations using white markers. \textit{Right,} GCs theorised to be associated with GSE from the study of \citet{myeong2019} are shown in black. 
 }
	\label{fig:cloudplot}
\end{figure*}

\subsection{Possible Constraints from Globular Clusters}
\label{sec:gsegcs}

In Fig.~\ref{fig:cloudplot}, we again plot the accreted and in-situ field star populations, but this time present the GALAH sample as a column-normalised ``cloud'', binned first in metallicity and then column-normalised. In addition to the field star populations, we include, in the middle panel, all the GCs in our sample which have been classified as in-situ by \citet{belokurov2023b} (after removing NGC~5139 and NGC~6273 which they note are likely mis-classified) and in the final panel, GCs which were flagged as being associated with GSE by \citet{myeong2019}. 

To bring our GC abundances onto the same scale as GALAH, we compared the [Si/Fe] and [Eu/Fe] abundances for clusters in common between our study and GALAH DR3 (NGC~288, NGC~362, NGC~1851 and NGC~5139). Note that we choose not to adopt GALAH abundances for any of the GCs in our sample as they are known to exhibit unphysical chemical trends with stellar parameters. This will be addressed in future abundance releases of GCs from GALAH. We calculated the offsets using only RGB stars in the GALAH GC sample in an effort to match our high resolution compilation methodology as closely as possible. On average we found an offset of $\sim+0.15$~dex in [Si/Fe] between our GC compilation and that of GALAH and and offset of $\sim-0.05$ in [Eu/Fe]. As such, the GCs were shifted downwards by 0.15~dex in [Si/Fe] and upwards by 0.05~dex in [Eu/Fe] (resulting in an upwards shift of 0.20~dex in [Eu/Si]) in Fig.~\ref{fig:cloudplot}.

\begin{figure*}
    \centering
	\includegraphics[scale=0.7]{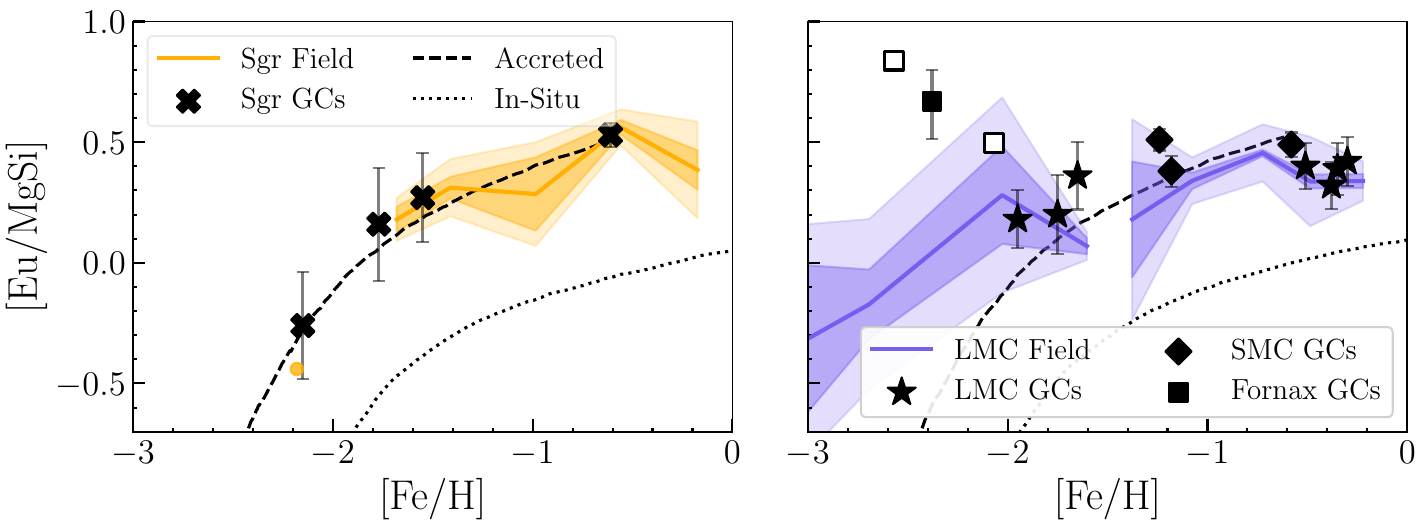}
    \caption{\textit{Left,} binned mean abundance values of Sgr field stars from \citet{reichart2020} in [Eu/Mg] vs. [Fe/H] are shown in yellow. The median absolute deviation weighted by square root the number of stars per bin is shown as the dark shaded region and 1.4826 times the median absolute deviation  (approximately one standard deviation) as the lighter shaded region. GCs associated with Sgr from the study of \citet{tango} are  over-plotted with black crosses (using their abundance of [Eu/Si] listed in Table~\ref{tab:litcomp}). Note the overlap between the GCs and field stars. \textit{Right,} the binned mean abundance and dispersion for LMC field stars from the studies of \citet{oh2023}, \citet{reggiani2021}, \citet{vanderbarlmc} and \citet{pompeialmcdisc} is shown in purple. LMC GCs from the studies of \citet{mucciarellilmc1} and \citet{mucciarellilmc2} are overplotted as stars. SMC and Fornax GCs are taken from spectroscopy of resolved GC stars from the studies of \citet{mucciarellismc} and \citet{letarte2006} respectively. Chemical evolution models tailored \textit{individually} to GSE (accreted) and the MW (in-situ) are plotted in both panels as the dashed black line and dotted black line, respectively to guide the eye. The details of the models are discussed in Sec.~\ref{sec:gcemodels}.}
    \label{fig:eusilmcsgr}
\end{figure*}

Beginning first with the final (right-most) column, of Fig.~\ref{fig:cloudplot} we show the intersection of our GC data set with the population of GSE GCs published in \citet{myeong2019}. A KS test of the \citet{myeong2019} GSE-tagged GCs and the in-situ GALAH, RPA, MINCE samples returns a $p$-value of 0.01, suggesting the GSE GCs were not drawn from the in-situ field star distribution. Comparing \citet{myeong2019} GSE GCs to the mean accreted RPA, MINCE and GALAH samples returns a $p$-value of 0.33, suggesting they could have come from the same distribution.
We find the agreement between the GSE GCs and the accreted field star population across metallicities, in all three panels, remarkable for two reasons. 

The first reason is how tight the sequence of GCs is across the three ratios. We speculate that this may be the clearest rise in [Eu/$\alpha$] seen in any dataset to-date, potentially providing the strongest support for the importance of NS mergers at early times. As discussed in the previous section, the prediction of large dispersion in [Eu/Fe] (and by extension [Eu/Si]) at low-metallicities by hydrodynamic simulations is not seen in the population of GSE GCs. In this way, the elemental abundances of GCs could be used to place constraints on the chemical history of the progenitor down to low metallicities with higher precision that is currently achieved in field stars. We revisit the validity of this statement in the next section. 

The second reason we find the agreement between GSE field stars and GCs to be remarkable, is the possibility of adding an additional dimension through the relatively precise GC ages. At low-metallicities sub-Gyr age precision from isochronal ages are possible for GCs \citep{ying2023}. In contrast, while astroseismic ages of individual field stars stars below $\mathrm{[Fe/H]}\sim-2$ may be possible \citep[as demonstrated in][]{puls2022}, uncertainties increase drastically beyond five Gyr (from $\pm1$~Gyr at 9~Gyr, to $\pm3$~Gyr at 13~Gyr). Label-transfer methods for field stars trained on high quality astroseismic ages also show large discrepancies for stars older than 10~Gyr \citep[ages are underpredicted by $\sim3$~Gyr on average][]{mackereth2019}.

In the case of datasets like \citet{vandenberg2013} or \citet{marinfrench2009}, large numbers of GC ages ($>50$) derived using a consistent methodology (choice of isochrone) can provide ages down to metallicities of $\mathrm{[Fe/H]}\sim-2$ with a precision perhaps unachievable in field stars. In this way, the GCs can provide precise timestamps for the GCE model, mapping between metallicity and time. In the case of the GSE GCs, cross-matching the GSE-tagged GCs in our sample using the list of \citet{myeong2019} with ages from \citet{vandenberg2013} results in the ratio of [Eu/Si] increasing from 0 to $\sim0.5$~dex over a period of one Gyr. The entire sequence from $\mathrm{[Eu/Si]}\sim-0.25$~dex to $\mathrm{[Eu/Si]}\sim0.5$~dex spans 1.75~Gyr, placing constraints on the star formation efficiency and history of the GSE progenitor.

Moving to the in-situ GCs (shown in the middle panel of Fig.~\ref{fig:cloudplot}), we can see that at metallicities above [Fe/H]$\sim-1.5$, the in-situ group follows the in-situ field stars fairly nicely in [Si/Fe]. In [Eu/Fe], the in-situ GCs display increased dispersion, spanning both the in-situ and accreted populations, but they show a clear downturn above metallicities of $\mathrm{[Fe/H]}>-0.5$, in-line with the in-situ population. In [Eu/Si] however, the high-metallicity GCs show on average better agreement with the in-situ field stars on average, aided by the distinguishing power of [Si/Fe]. 

A KS-test comparing the mean [Eu/Si] values of the in-situ GCs across metallicities to the mean in-situ RPA, MINCE and GALAH samples cannot reject the hypothesis that the they were drawn from the same distribution ($p=0.17$). Comparing the in-situ GCs to the mean accreted field star sample returns a $p$-value of $p=0.03$, strongly suggesting they we not drawn from the same distribution.

At low-metallicities ($\mathrm{[Fe/H]}<-1.50$), the agreement between in-situ field stars and GCs breaks down. Of the GCs in this region, NGC~6397 ($\mathrm{[Fe/H]}$=-1.99) and NGC~6205 ($\mathrm{[Fe/H]}$=-1.50) have the second and third highest values of [Eu/Si] ($\sim0.26-0.29$~dex). In the case of NGC~6205, it is the closest GC to the accreted/in-situ boundary in E-L$_{\mathrm{z}}$. For this reason we suggest it may be mis-classified and have marked it with a black dot inside the white marker in all panels. The high [Eu/Si] value of NGC~6397 is less easy to explain, as Eu is genuinely high in the GC.

The other major outlier in [Eu/Si] among the in-situ GCs, is the lowest metallicity GC, NGC~7078.  We have also marked this cluster with a black dot in the middle panel of Fig.~\ref{fig:cloudplot}. This enhancement in [Eu/Si] in NGC~7078 is at a level unseen among accreted field stars at the same metallicity. This is driven by a large, but uncertain Eu-enhancement ($\mathrm{[Eu/Fe]}=0.8\pm0.25$). Recently, \citet{Cabrera24} confirmed a large spread in [Eu/Fe] ($\sigma\sim0.2$) in the cluster using measurements of 62 stars \citep[significantly more than the study of][]{ngc7078ref}. Note that \citet{Cabrera24} do not report a value [Si/Fe] for the cluster, hence why we have adopted [Eu/Si] from \citet{ngc7078ref}.

\citet{Cabrera24} report a similar mean value of [Eu/Fe] to \citet{ngc7078ref} ($\mathrm{[Eu/Fe]}\sim0.7$) but note that this is the average of two distinct populations of Eu-stars \citep[also observed in][]{Worley2013}. They detect a possible correlation between Na and Eu, contrary to what \citet{roederer11} found. However, they hypothesise that this is the result of $r$-process enrichment during the formation of the first generation of stars - maintaining a connection between the host environment. We speculate that the large overall Eu-enhancement of the cluster is likely linked to the proximity of the $r$-process enrichment event.

\section{Globular Clusters Trace the Chemical Histories of their Hosts}
\label{sec:globhist}
In this section, we continue to explore the agreement between field star abundances in dGals and their clusters, across metallicities. Motivated by the results in Sec.~\ref{sec:gsegcs}, we focus on other dGal systems, specifically the Large M. Cloud\footnote{We acknowledge that the continued use of the name Magellan is both potentially traumatic for Indigenous peoples and factually incorrect, as he did not discover the clouds. We advocate for the adoption of a new naming scheme and abbreviate the name only for the remainder of this paper.} (LMC) and Sagittarius (Sgr) dGal and their clusters. We focus on these two galaxies because they i) have large systems of GCs with high quality measurements, and ii) have published field star [Eu/Si] abundances across a large range of metallicites. Pushing beyond the MW, and thanks to the groundbreaking technique of extracting high-precision abundances from integrated light spectroscopy \citep{sakarim31}, we explore the [Eu/Si] abundances in GCs around M~31 and discuss the potential of classifying extra-galactic GCs as in-situ and accreted.

\subsection{In Local Group dGals}
Fig.~\ref{fig:eusilmcsgr} presents data for field stars and clusters in Sgr and the LMC, showcasing the remarkable agreement between the evolution of [Eu/Si] as a function of metallicity across \textit{both} field stars \textit{and} clusters in the two systems. Sgr is shown in the left panel, where the mean abundance (and dispersion) of [Eu/Mg] is plotted using field stars from the study of \citet{reichart2020} in yellow. Sgr GCs identified (or confirmed) by \citet{tango} are included using the black crosses taking their [Eu/Si] abundances from our literature compilation.

The right panel shows the mean values (and dispersion) of [Eu/Si] in LMC field stars from \citet{pompeialmcdisc}, \citet{vanderbarlmc}, \citet{reggiani2021} and \citet{oh2023} in purple, with LMC GCs from the studies of \citet{mucciarellilmc1} and \citet{mucciarellilmc2} over-plotted with star symbols. Abundances from \citet{mucciarellilmc1} and \citet{mucciarellilmc2}, are derived from between five to eleven stars in each GC. Note that the extremely low-metallicity data from \citet{oh2023} includes upper limits for five of the seven stars in their study. 

Due to a lack of Eu-abundances for SMC field stars, we include SMC GCs from the recent study of \citet{mucciarellismc} alongside the LMC data in Fig.~\ref{fig:eusilmcsgr}. The SMC GCs include the GCs NGC~121, NGC~339 and NGC~419 using measurements of between five to eight stars in each. Due to the large metallicity difference between Fornax GCs and Fornax field stars with Eu measurements from the studies \citet{letarte2006} and \citet{letarte2010}, respectively (0.5~dex), we also include Fornax GCs in the panel showing data for the LMC. Data for the Fornax GCs, 1, 2 and 3 is taken from the study of \citet{letarte2006} averaging measurements for two to three stars in each.

Because \citet{letarte2006} did not measure Si in their study, we adopt Ca as our $\alpha$-tracer to include the Fornax GCs. We justify this choice by noting that the published Ca abundances of GCs in our compilation are on average only 0.1~dex lower than Si. Note that the Fornax field star data truncated at $\mathrm{[Fe/H]}=-1.5$ show the same plateau at [Eu/Ca] = 0.7 as the GCs. The Fornax GCs appear super enhanced in [Eu/Ca] at very low metallicities, which is driven in the case of Fornax 3 (the only GC with a true measurement, rather than an upper limit, denoted by open symbols in Fig.~\ref{fig:eusiexternal}), by truly enhanced Eu ($\sim0.90\pm0.10$~dex).

To provide further support for the idea that GCs trace the star formation history of their hosts, we include predictions from the individually-fit GCE models introduced in the previous section. The dGal GCE model tailored to GSE is plotted using the dashed line, while the MW model is plotted with a dotted line in both panels. The agreement between our GCE model tailored to GSE and the Sgr stars (both field and GCs) is striking. Although age is not included in these plots, the GCE models reach the metallicity of the 7-8~Gyr old cluster Terzan~5 in Sgr but do not achieve the metallicity of the much younger (1-2~Gyr) LMC GCs. In our joint GSE-MW GCE model, the GSE maximum age is ($6\pm3$) Gyr explaining this difference. 

In their study of star formation histories in dGals, \citet{Hasselquist2021} derive a star formation efficiency in GSE that is five times higher than the efficiency in Sgr. Therefore, we interpret the fit to the Sgr data as further evidence that SFE alone cannot explain the appearance of elevated [Eu/Si] in dGals at late times.

\subsection{Beyond the MW, in M31}
We now expand our exploration of [Eu/Si] beyond the MW into the system of GCs around M~31.
Unfortunately, Eu measurements have not been made for field stars in M~31, though we note that $\alpha$-element abundances have been made in resolved field stars in M~31 \citep{escala2019}. Therefore, we utilise our MW in-situ and accreted field star abundances for comparison in this section.

Data for GCs in the halo of M~31 comes from the integrated light study of \citet{sakarim31}, who examined GCs associated with substructure around the galaxy. These measurements include data for the GCs, PA~17, H~23, H~10, PA~53, PA~56 and PA~06, three of which have only upper limits of Eu. Note that the naming conventions for these GCs follows that of \cite{sakarim31} which in-turn adopts the conventions of \citet{huxor2008} and \citet{huxor2014}. Note that \citet{sakarim31} uses the the 6645~\AA\, line of Eu is to determine Eu-abundances, which is consistent with our GC compilation. Upper limits for denoted using empty symbols, while measurements are filled (and have corresponding uncertainties).

\begin{figure}
    \centering
	\includegraphics[width=\linewidth]{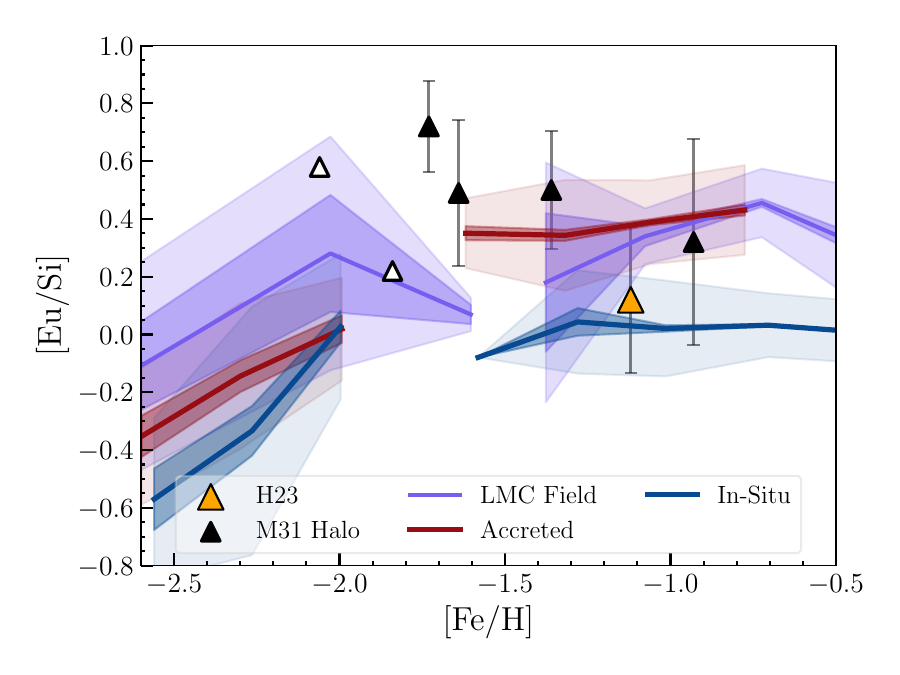}
    \caption{The distribution of [Eu/Si] as a function of metallicity for GCs in external galaxies. LMC, MW (in-situ) and GSE (accreted) field stars means are shown alongside the median absolute deviation weighted by square root the number of stars per bin (dark shaded region), and 1.4826 times the median absolute deviation as (lighter shaded region). M~31 GC abundances are derived from analysis of their integrated light from \citet{sakarim31}. Open symbols denote upper-limits. Note the overabundance of [Eu/Si] in the GCs associated with substructure around M~31. The M~31 GC H~23, the closest to the main body of M~31 and unassociated with substructure, is marked in orange; to highlight its proximity to the MW GCE model supporting an in-situ origin.}
    \label{fig:eusiexternal}
\end{figure}

The distribution of the M~31 GCs in [Eu/Si] vs. [Fe/H]-space is shown in Fig.~\ref{fig:eusiexternal}, alongside the mean LMC field star and GALAH/MINCE/RPA in-situ and accreted sample abundances. The first thing of note is the continued appearance of an elevated ratio of [Eu/Si] in the GCs associated with the halo of M~31 relative to the MW in-situ field star sample, with the exception of the M~31 GC, H~23. This enhancement mirrors what is seen in the population of GCs associated with GSE (e.g. see the lower right panel in Fig.~\ref{fig:cloudplot}).  

An explanation for the single outlying GC, H~23, is an interesting and potentially powerful one. When considering the projected distance of the GCs around M~31, H~23 lies the closest to the main body of M~31. H~23 is also the second most metal-rich GC ($\mathrm{[Fe/H]}=-1.12$) in the M~31 sample and was found by \citet{sakarim31} to have chemistry consistent with both MW field stars and MW GCs at the same metallicity. Finally, \citet{sakarim31} also note that despite initial evidence that H~23 could be associated with Stream D in M~31 from the study of \citet{mackey2010}, that follow-up radial velocity measurements from \citet{veljanoski2014} likely rule this out.

Given that the chemo-dynamical evidence surrounding H~23 suggests that it may not be associated with substructure, it could be that its association with the GCE track of a larger galaxy suggests an in-situ origin. If this is the case, and the trend of an overabundance of [Eu/Si] in other dGal GCs is to be believed, the ratio of [Eu/Si] in GCs surrounding other galaxies could be used to discern which originated around dGals and were then later accreted. In the era of extremely large telescopes, with the ability to resolve large numbers of stars in GCs around other galaxies, this could be a means of exploring the accretion history of those galaxies \citep[e.g. using ANDES on the Extrelemy Large Telescope,][]{andes}.

\section{Conclusions}
\label{sec:conclusions}
In this study we explored the use of the $r$-process element, europium (Eu) as a chemical differentiator between populations of in-situ and accreted field stars and their globular clusters (GCs). By combining information from light ($\alpha$) and heavy ($r$-process) elements, we find the ratio of [Eu/Si] resolves the different star formation histories of both in-situ and accreted field stars and GCs across a large range in metallicities ($-2.6\leq\mathrm{[Fe/H]}\leq0$), with more [Eu/Si] measurements needed at the lowest-metallicities to confirm this.

The main results of this study are summarised in the following:

\begin{itemize} 
    \item We demonstrate the existence of a significant change in the overall abundance ratio of [Eu/Si] across the dynamical boundary in E-L$_{\mathrm{z}}$ separating in-situ from accreted stars \cite[as defined previously by, ][]{Myeong2018action, belokurov2023a}. Stars with energies higher than approximately Solar (accreted) exhibit an elevated value of [Eu/Si], largely driven by [Eu/Fe], while those below this energy level (in-situ) have lower [Eu/Si]. At high metallicities, these differences persist through two distinct plateaus across metallicity in the two populations.
    \item We provide evidence that these global [Eu/Si] trends may persist down to low metallicities ($\mathrm{[Fe/H]_{NLTE}}=-2.5$). While the average levels of [Eu/Si] decrease with decreasing [Fe/H] across both samples, in-situ stars with lower total energies have systematically lower [Eu/Si] relative to accreted stars above the dynamical boundary. This implies that the [Eu/$\alpha$] scatter within a single galaxy is lower than previously estimated.
    \item We suggest that the difference between the two sequences is not caused by differences in star formation efficiency (SFE) as previously suggested \citep{matsuno21, naidu2021}, with the caveat that we are only fitting data above $\mathrm{[Fe/H]}\geq-1.8$. This is supported by the failure of SFEs derived from high-quality fits to the $\alpha$-[Fe/H] plane to fit either [Eu/Fe] or [Eu/Si]. Follow-up multi-zone GCE models will explore whether a multi-phase ISM is needed to explain the differences.
    \item Because of this difference, we suggest galactic chemical evolution models and hydrodynamic simulations consider the origin (accreted/in-situ) of stars in when comparing to observational data sets. To differentiate the two populations, the simple dynamical cut in $E$-$L_{z}$ introduced in \citet{belokurov2023a} could be applied. 
    \item From a literature compilation of 54 globular clusters (GCs), we find that accreted GCs trace the [Eu/Si] abundance evolution of accreted field stars across metallicities remarkably well. These results support the primordial origin of Eu in some GCs, as suggested by previous works \citep{roederer11, kochhansen2021, monty2023a, kirby2023}. 
    \item Using ages from \citet{vandenberg2013} for the population of GSE GCs taken from \citet{myeong2019}, an enrichment of 0.75 dex in [Eu/Si] (from -0.25 to 0.5~dex) occurs in GSE over the time span of 1.75~Gyr, presenting the opportunity for sub-Gyr timing resolution when building host galaxy star formation histories.
    \item In both accreted and in-situ field stars \textit{and} more significantly in GCs, we find a downwards trend in [Eu/Si] at low-metallicities. This trend persists after applying NLTE corrections to Eu. We suggest that this preferentially supports enhancement by neutron star mergers over magneto-rotational supernovae (which produce a flat trend in [Eu/Si]).
    \item We suggest that [Eu/Si] could be used as a chemical tag beyond the MW, based on evidence that GCs trace field star [Eu/Si] abundances in Local Group dwarf galaxies (dGals). We show this is also true in the halo of M~31, where the GC H~23 displays an in-situ-like [Eu/Si] abundance ratio which supports its lack of association with substructure.
    
\end{itemize}

\section*{Acknowledgements}
We thank the referee for their helpful comments and suggestions. SM wishes to thank Erika Holmbeck for her incredibly helpful comments that greatly improved the paper. SM also wishes to thank Eugenio Carretta, Angela Bragaglia, Kim Venn, Keith Hawkins, Tadafumi Matsuno and Zhen Yuan for helpful conversations on abundances in metal-poor stars and GCs and the organisers of the 2023 MIAPbP meeting ``Stellar Astrophysics in the Era of Gaia, Spectroscopic, and Astroseismic Surveys'', Maria Bergemann, Daniel Huber, Saskia Hekker, Amanda Karakas and Rolf-Peter Kudritzki for creating an incredibly enriching and collaborative environment where this work began. JLS acknowledges support from the Royal Society (URF\textbackslash R1\textbackslash191555). T.T.H acknowledges support from the Swedish Research Council (VR 2021-05556). AAA acknowledges support from the Herchel Smith Fellowship at the University of Cambridge and a Fitzwilliam College research fellowship supported by the Isaac Newton Trust. Based on observations made with ESO Telescopes at the La Silla Paranal Observatory under programme IDs 072.D-507 and 073.D-0211. This work relies heavily on the \texttt{Astropy} \citep{astropy1, astropy2}, \texttt{SciPy} \citep{scipy}, \texttt{NumPy} \citep{numpy} and \texttt{Matplotlib} \citep{matplotlib} libraries and \texttt{Jupyter} notebooks \citep{jupyter}. This work makes use of data from the GALAH survey, observed using the Anglo-Australian Telescope on the unceded territory of the Gamilaraay people. We acknowledge the traditional custodians of the land and pay our respects to elders past and present.

\section*{Data Availability}
This study makes use of existing literature data as shown in Table~\ref{tab:litcomp} and public data releases. Equivalent widths for the two GCs analysed in this study are available upon request. 



\bibliographystyle{mnras}
\bibliography{bib} 




\appendix

\section{Abundance Analysis of NGC~1851, NGC~1904 and NGC~2808}
\label{app:chemabundanceanalysis}
As these clusters have been studied extensively in the literature, we did not re-derive the abundances for the full extent of elements measurable in the UVES spectra. Instead, we only re-measured the abundance of Mg, Si and Ca in the clusters and the abundance of Eu. To do this, we measured equivalent widths and performed a classical spectroscopic analysis, assuming 1D Local Thermodynamic Equilibrium (LTE) utilising plane parallel atmospheres. As these techniques have been described extensively in the literature \citep[including in,][]{monty2023a}, we avoid re-hashing the process step-wise here and instead provide a brief description of the process. 

Briefly, equivalent widths were measured for the 5528~\AA\ and 5711~\AA\ \ion{Mg}{I} lines line using both \textsc{IRAF's} \texttt{splot} task and the automated line measurement software \texttt{DAOSpec} \citep{daospec}. Measurements between \texttt{DAOSpec} and \texttt{splot} were found to be extremely consistent in the region of $10<\mathrm{EW}<70$~m\AA, so EWs were determined for Si and Ca using \texttt{DAOSpec} alone. In total, nine lines of Si and 22 lines of Ca were selected \citep[from the linelist of, ][]{monty2023a} to determine the abundances of these two elements. All line measurements were culled to remove $>3\sigma$ outliers prior to abundance determination. 

The 6645~\AA\ line of \ion{Eu}{II} was synthesised to determine the abundance of Eu in all program stars. Hyperfine structure corrections were adopted from \citet{lawler2001}, using the \texttt{linemake} program \citet{linemake}. Isotopic splitting across the two Eu isotopes, 151 and 153, was also considered, adopting a fractional contribution of 53\% and 47\% respectively \citep{sneden08}. A macroturbulent velocity of 7~km/s was included for all of the stars, reflecting an average estimate of macroturbulence in evolved giants \citep{gray86, carney08}. Finally, the synthetic spectra were smoothed to reflect the average spectral resolution of the observations ($\sim47,000$) using a Gaussian function, full width at half maximum equal to 6645~\AA/47,000 = 0.14~\AA.

\begin{figure}
    \centering
	\includegraphics[scale=0.5]{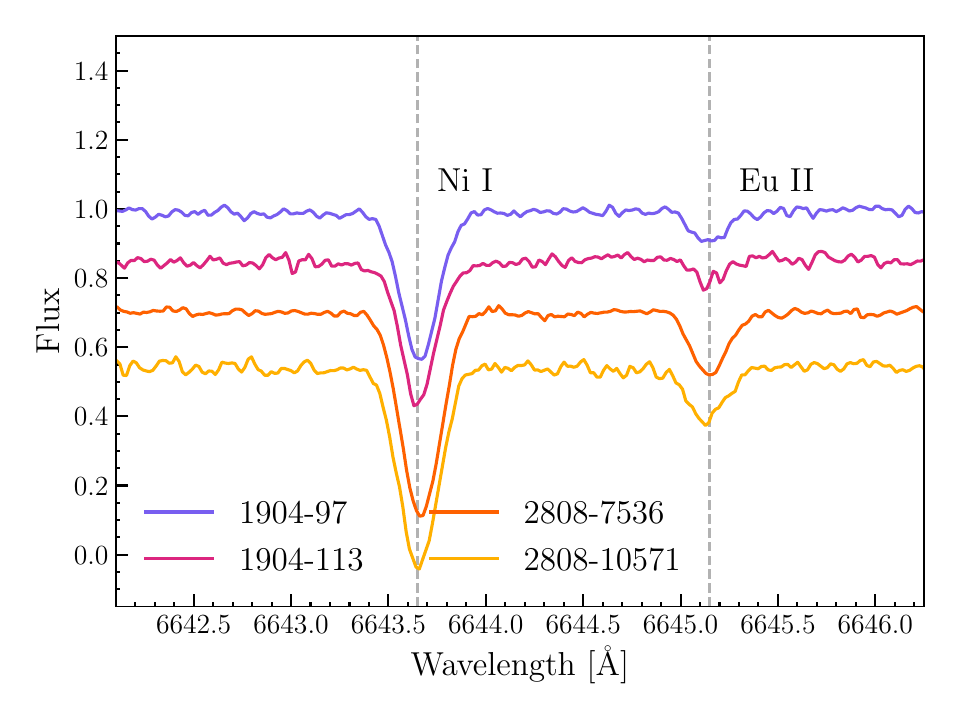}
    \caption{Example spectra highlighting the \ion{Eu}{II} line at 6645\AA\ in the highest SNR stars (1904-97, 2808-7536) and lowest SNR stars (1904-113 and 2808-10571) in the two clusters. A neighboring \ion{Ni}{I} line is also labeled.}
    \label{fig:eulines}
\end{figure}

Examples of the \ion{Eu}{II} line for the highest and lowest SNR stars in each cluster are shown in Fig.~\ref{fig:eulines}. The average of the two set of independent measurements were then adopted for further analysis. To determine the spectroscopic stellar parameters for the stars (effective temperature: $\mathrm{T}_{\mathrm{eff}}$, log surface gravity: log~$g$, metallicity: [Fe/H] and microturbulence: $\xi$), $~180$ \ion{Fe}{I} and $~15$ \ion{Fe}{II} lines from the linelist of \citet{monty2023a} were measured using \texttt{DAOSpec}. The \ion{Fe}{I}/\ion{Fe}{II} linelist was cleaned to remove obvious ($>3\sigma$ outliers) prior to determination of the stellar parameters. 

In the case of NGC~2808, the spectroscopic stellar parameters (SPs) were determined using the classical technique of minimising abundance trends with excitation potential and reduced equivalent width and balancing of the Fe ionisation states. All SP determinations and abundance measurements were performed using the \textsc{python}-based tool \texttt{q2} \citep{q2} to communicate to the radiative transfer code \texttt{MOOG} \citep{moog}. \texttt{MARCS} model atmospheres \citep{marcs} were adopted throughout. The final spectroscopic SPs and literature photometric SPs are listed in Table \ref{tab:sps} alongside those published in \citet{carretta2009b} (in the case of NGC~2808 and NGC~1904) and \citet{carretta1851} (in the case of NGC~1851). The uncertainties we quote on the stellar parameters are not absolute and instead reflect the precision of the minimisation of abundance trends and FeI vs FeII abundance differences (standard 1D spectroscopic determination of SPs).

Given the metallicity of NGC~2808, and the agreement between the spectroscopic and photometric SPs and their respective abundances, we adopt the abundances determined using the spectroscopic parameters for the remainder of this study. In the case of the lower-metallicity GC NGC~1904 \citep[$\mathrm{[Fe/H]}\approx-1.6$][]{carretta2009b}, we adopted the photometric SPs published in \citet{carretta2009b} to avoid issues associated with spectroscopically determined SPs at low-metallicities \citep{mucciarelli2020}. 

When adopting the photometric SPs for the stars in NGC~1904, the resultant Mg-abundances were significantly lower than those published in \citet{carretta2009b}. In the most extreme case, the Mg abundances differed by $\sim0.4$~dex. This is likely because the broad Mg lines are very sensitive to the choice effective temperature. Both the Eu and Si abundances were observed to change by no more than $\sim0.1$~dex (on the order the measurement uncertainty). This is likely because the Si and Eu lines are significantly weaker than the Mg lines, making them less sensitive to the choice of SPs. When using the optimal spectroscopic SPs derived for NGC~1904 (using the classical spectroscopic optimisation technique), the Mg abundances were brought into much better agreement, differing by no more than $0.19$~dex. Ultimately, because of the disagreement in Mg abundances observed when adopting spectroscopic or photometric SPs, and the concerns discussed in Sec.~\ref{sec:litdat}, we focus on the recovery and verification of Si abundances between our study and that of \citet{carretta2009b} and adopt Si as our $\alpha$-element tracer moving forward.

The abundance determinations for Mg, Si, Ca and Eu are shown in Table~\ref{tab:abunds} and compared (in the case of Mg and Si) to those determined by \citet{carretta2009b} and \citet{carretta1851}. The abundances are solar-scaled using the Solar abundances of \citet{asplund09}. Note that \citet{carretta2009b} do not adopt the same solar scale, as they perform differential line-by-line abundance determination relative to the Sun. However, we do not expect the offsets to be larger than the abundance uncertainties. In general, we find good agreement with \citet{carretta2009b, carretta1851} in  Mg and Si, with 9/12 stars displaying consistent (within measurement uncertainties) across the three studies. In the case of our calibration cluster, NGC~1851, the values of Eu are remarkably consistent (with the exception of NGC~1851-26271, for which \citet{carretta1851} find a higher Eu abundance). The values of [Mg/Fe] are the most disparate across studies, in addition to having the largest uncertainties. We take this as evidence to support our choice of an alternative $\alpha$-element tracer. However, we note that the disagreement could be be driven by our choice to fit all lines with Gaussian profiles, when the 5528~\AA\, line of Mg may require fitting with a Voigt profile. The cluster averages are derived by combining the second to last column as a weighted average, the cluster standard deviation is also quoted. 

\begin{table*}
	\centering
	\caption{Compilation of stellar parameters (effective temperature: $\mathrm{T}_{\mathrm{eff}}$, log surface gravity: log~$g$, metallicity: [Fe/H] and microturbulence: $\xi$) derived in this study and from the study of \citet{carretta2009b} for stars in NGC~1904 and NGC~2808 and \citet{carretta1851} for stars in NGC~1851. The optimal spectroscopic SPs derived using the method described in Sec.~\ref{app:chemabundanceanalysis} are listed first, followed by the photometric SPs published in \citet{carretta2009b} and \citet{carretta1851}.}
	\label{tab:sps}
	\begin{tabular}{lllllllll} 
		\hline
        & \multirow{1}{*}{Spectro.} & & & & \multirow{1}{*}{Photo.} & & & \\
        \hline
		Star & $\mathrm{T}_{\mathrm{eff}}$ & log~$g$ & [Fe/H] & $\xi$ & $\mathrm{T}_{\mathrm{eff}}$ & log~$g$ & [Fe/H] & $\xi$\\
        ... & [K] & [dex] & [dex] & [cm/s$^{2}$] & [K] & [dex] & [dex] & [cm/s$^{2}$] \\
		\hline
		NGC1904-97 & $4479\pm18$ & $0.89\pm0.13$ & $-1.69\pm0.03$ & $1.52\pm0.03$ & 4378 & 1.23 & -1.62 & 1.49\\
		NGC1904-98 & $4483\pm31$ & $1.04\pm0.10$ & $-1.69\pm0.04$ & $1.44\pm0.04$ & 4386 & 1.25 & -1.57 & 1.92\\
		NGC1904-149 & $4609\pm29$ & $1.16\pm0.11$ & $-1.62\pm0.04$ & $1.39\pm0.04$ & 4512 & 1.51 & -1.62 & 1.77\\
        NGC1904-113 & $4561\pm33$ & $1.10\pm0.11$ & $-1.61\pm0.05$ & $1.49\pm0.05$ & 4430 & 1.34 & -1.54 & 1.66\\
        \hline
        NGC2808-7536 & $4283\pm26$ & $0.86\pm0.17$ & $-1.18\pm0.03$ & $1.51\pm0.03$ & 4311 & 1.19 & -0.98 & 1.69\\
		NGC2808-38660 & $4318\pm37$ & $1.12\pm0.12$ & $-1.12\pm0.04$ & $1.54\pm0.04$ & 4322 & 1.22 & -1.21 & 1.62\\
		NGC2808-8603 & $4365\pm29$ & $1.19\pm0.12$ & $-1.16\pm0.03$ & $1.51\pm0.04$ & 4343 & 1.24 & -1.18 & 1.66\\
        NGC2808-10571 & $4334\pm30$ & $1.20\pm0.12$ & $-1.12\pm0.03$ & $1.55\pm0.03$ & 4315 & 1.21 & -1.21 & 1.66\\
        \hline
        NGC1851-26271 & $3986\pm31$ & $0.50\pm0.13$ & $-1.17\pm0.04$ & $1.81\pm0.04$ & 3910 & 0.54 & -1.24 & 2.13\\
		NGC1851-28520 & $4273\pm50$ & $1.04\pm0.11$ & $-1.15\pm0.04$ & $1.62\pm0.05$ & 4141 & 1.01 & -1.27 & 1.83\\
		NGC1851-32903 & $4058\pm43$ & $0.67\pm0.16$ & $-1.10\pm0.05$ & $1.64\pm0.05$ & 4040 & 0.80 & -1.06 & 1.66\\
        NGC1851-39801 & $4099\pm39$ & $0.64\pm0.14$ & $-1.16\pm0.04$ & $1.69\pm0.04$ & 3979 & 0.68 & -1.19 & 1.83\\
		\hline
	\end{tabular}
\end{table*}

\begin{table*}
    \centering
    \caption{Abundances of Mg, Si and Ca and Eu determined using spectroscopic SPs determined as described in Sec.~\ref{app:chemabundanceanalysis}. Errors associated with single line measurements (Eu) are a result of the uncertainty in the SPs remaining uncertainties quote the abundance dispersion arising from the set of line measurements. After each abundance we list the abundance difference between our study and that of \citet{carretta2009b} (in the case of NGC~1904 and NGC~2808) or \citet{carretta1851} (in the case of NGC~1851), the difference is defined as $\Delta\mathrm{[X/Fe]}=$~\citet{carretta2009b, carretta1851} - this study. The uncertainty of the associated the abundance measurement from \citet{carretta2009b, carretta1851} is listed in brackets after each difference.}
	\label{tab:abunds}
    \begin{tabular}{llllllll} 
		\hline
    	Star & [Mg/Fe] & $\Delta$ [Mg/Fe] & [Si/Fe] & $\Delta$ [Si/Fe] & [Ca/Fe] & [Eu/Fe] & $\Delta$ [Eu/Fe] \\
	\hline
		NGC1904-97 & $0.33\pm0.06$ & 0.00 (0.05) & $0.30\pm0.05$  & -0.03 (0.09) & $0.28\pm0.05$ & $0.47 \pm 0.06$ & ... \\
		NGC1904-98 & $0.31\pm0.06$ & 0.02 (0.13) & $0.33\pm0.05$  & -0.07 (0.06) & $0.17\pm0.05$ & $0.57 \pm 0.11$ & ...\\
		NGC1904-149 & $0.13\pm0.06$ & 0.20 (0.13) & $0.34\pm0.05$  & 0.00 (0.06) & $0.37\pm0.05$ & $0.30 \pm 0.11$ & ...\\
        NGC1904-113 & $0.08\pm0.06$ & 0.08 (0.05) & $0.32\pm0.05$  & 0.00 (0.07) & $0.26\pm0.05$ & $0.34 \pm 0.16$ & ... \\
        \hline
        NGC2808-7536 & $0.27\pm0.15$ & 0.04 (0.05) & $0.32\pm0.06$  & -0.08 (0.06) & $0.21\pm0.05$  & $0.58 \pm 0.10$ & ... \\
		NGC2808-38660 & $0.23\pm0.13$ & 0.17 (0.12) & $0.32\pm0.05$  & -0.09 (0.08) & $0.25\pm0.06$ & $0.70 \pm 0.06$ & ... \\
		NGC2808-8603 & $0.23\pm0.16$ & 0.04 (0.23) & $0.34\pm0.05$  & -0.03 (0.12) & $0.26\pm0.05$  & $0.89 \pm 0.06$ & ... \\
        NGC2808-10571 & $0.21\pm0.14$ & 0.15 (0.19) & $0.30\pm0.06$  & -0.02 (0.09) & $0.22\pm0.05$ & $0.68 \pm 0.06$ & ... \\
        \hline
        NGC1851-26271 & $0.22\pm0.13$ & 0.16 (0.21) & $0.32\pm0.05$  & 0.09 (0.06) & $0.12\pm0.06$  & $0.70 \pm 0.06$ & 0.14 (0.03)\\
		NGC1851-28520 & $0.19\pm0.14$ & 0.14 (0.19) & $0.27\pm0.06$  & 0.10 (0.12) & $0.20\pm0.06$  & $0.73 \pm 0.11$ & 0.06 (0.03) \\
		NGC1851-32903 & $0.27\pm0.20$ & 0.04 (0.19) & $0.34\pm0.08$  & 0.09 (0.11) & $0.20\pm0.07$  & $0.58 \pm 0.11$ & 0.10 (0) \\
        NGC1851-39801 & $0.24\pm0.16$ & 0.15 (0.04) & $0.30\pm0.05$  & 0.07 (0.12) & $0.17\pm0.06$  & $0.64 \pm 0.06$ & -0.06 (0) \\
		\hline
	\end{tabular}
\end{table*}

\section{Exploring the distribution of different GSE-identified GCs}
\label{app:gsecompilations}
\begin{figure*}
    \centering
	\includegraphics[scale=0.6]{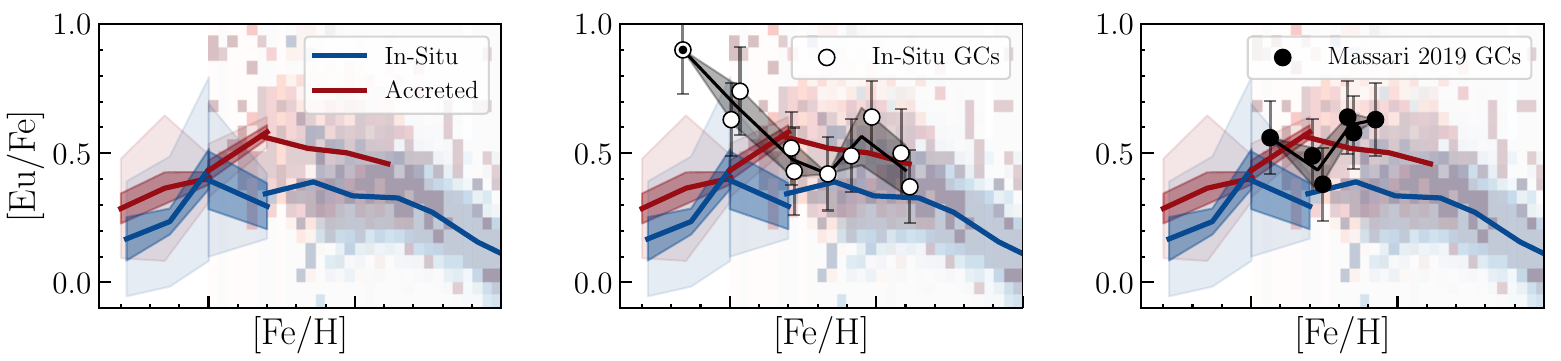}
    \caption{Data from the homogeneous study of \citet{schiappa2024} examining neutron capture abundances in GCs. Of the 16 GCs in their study with [Eu/Fe] measurements, we classify ten as in-situ using the designation of \citet{belokurov2023b} and six as belonging to GSE \citep[as identified by][]{massari2019}.}
    \label{fig:schlapgcs}
\end{figure*}

To explore if the tight trend of increasing [Eu/Si] found among the accreted GCs \citep[specifically those tagged as belonging to GSE by][]{myeong2019} is significant, ideally we would like to plot the evolution of a homogeneously measured set of GSE GCs. The recent study of \citet{schiappa2024} provides homogeneously measured [Eu/Fe] for ten in-situ GCs and six GSE GCs \citep[as identified by ][]{massari2019}. Their measurements are shown in Fig.~\ref{fig:schlapgcs}. Of the GSE GCs, they only have three with metallicity $\mathrm{[Fe/H]}<-1.5$ making it difficult to assess if the rise in [Eu/Fe] we see in Fig. ~\ref{fig:cloudplot} is present in their data as well. Examining their in-situ GCs, we again see overlap with both the accreted stars in [Eu/Fe] and in-situ stars. Once again NGC~7078 is an outlier among the in-situ GCs with very high [Eu/Fe].

\begin{figure*}
    \centering
	\includegraphics[scale=0.4]{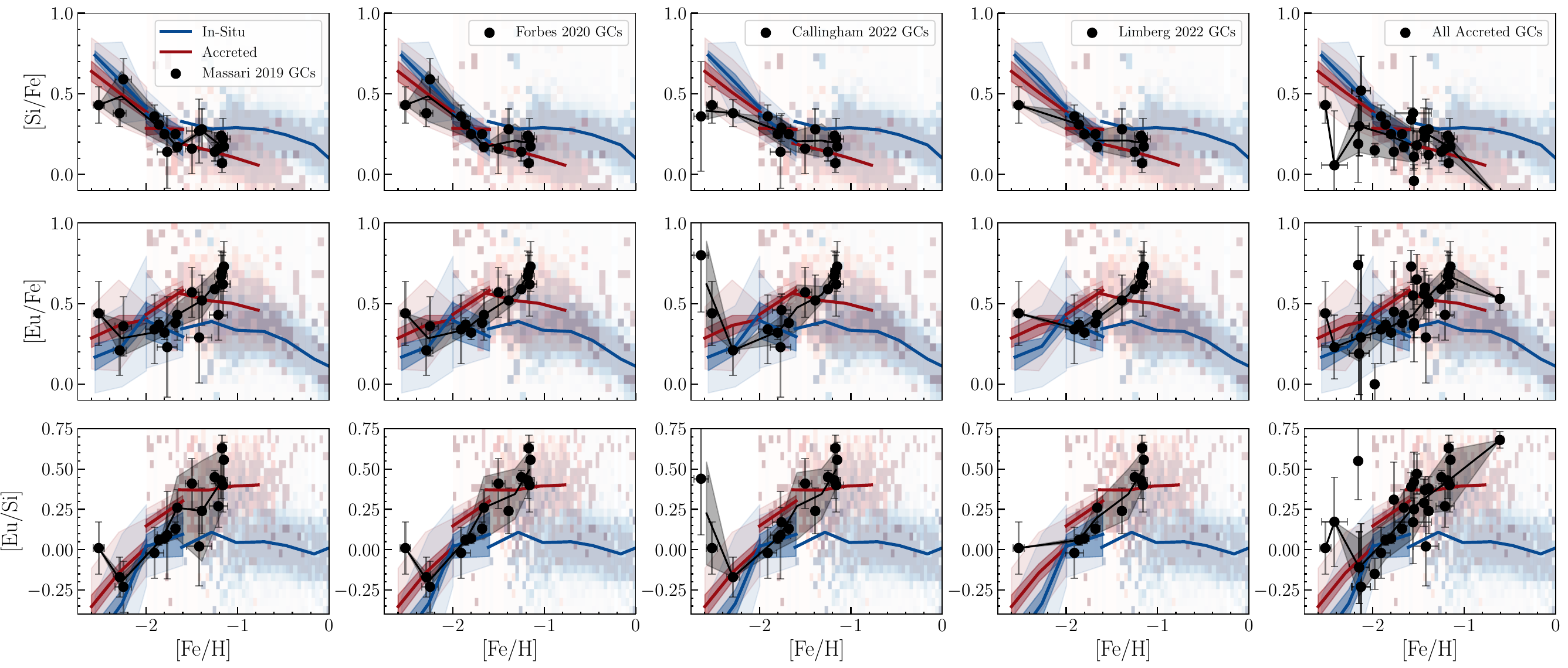}
    \caption{Same as Fig.~\ref{fig:cloudplot}, except the GSE-tagged GCs are taken from the studies of \citet{massari2019}, \citet{forbes2020}, \citet{callingham2022}, \citet{limberg2022} and \citet{malhan2022}, moving from left to right.}
    \label{fig:allgccombos}
\end{figure*}

In an attempt to further explore the robustness of the trend of increasing [Eu/Si] with metallicity. We considered classifications of GSE GCs from other studies, namely those of \citet{massari2019}, \citet{forbes2020}, \citet{callingham2022}, \citet{limberg2022} and \citet{malhan2022}. Using our compilation of abundances cross-matched with the GSE-tagged GCs from each study, we created Fig.~\ref{fig:allgccombos}. With the caveat that many of the studies identify the same GSE GC candidates, we can see that the trend present in Fig.~\ref{fig:cloudplot} is seen in every compilation. Although we cannot neglect the weakness of our compilation being inhomogeneous, it is reassuring to see that the trend persists.

\bsp	
\label{lastpage}
\end{document}